\documentclass[fleqn,usenatbib]{mnras}

\usepackage[T1]{fontenc}
\usepackage[utf8]{inputenc}
\usepackage{ae,aecompl}
\usepackage{siunitx}
\usepackage{enumerate}

\usepackage{graphicx}	
\usepackage{amsmath}	
\usepackage{amssymb}	
\usepackage{mathtools}
\usepackage{ulem}
\usepackage{xcolor}
\usepackage[multiple]{footmisc}
\usepackage{CJK}
\usepackage{newtxtext,newtxmath}
\title[FAST early pulsar discoveries: Effelsberg follow-up]{FAST early pulsar discoveries: Effelsberg follow-up}

\author[M. Cruces et al.]{
M.~Cruces$^{1\thanks{E-mail: mscruces@mpifr-bonn.mpg.de (MC)}, 2}$,
D.~J.~Champion$^{1}$,
D.~Li\begin{CJK}{UTF8}{bsmi}
(李菂)
\end{CJK}$^{2\thanks{E-mail: dili@nao.cas.cn},7,8}$,
M.~Kramer$^{1}$,
W.~W.~Zhu$^{2}$,
\newauthor
P.~Wang$^{2}$,
A.~D.~Cameron$^{3,4,5}$,
Y.T.~Chen$^{2}$,
G.~Hobbs$^{3}$,
P.~C.~C.~Freire$^{1}$,
E.~Graikou$^{1}$,
\newauthor
M.~Krco$^{2}$,
Z.~J.~Liu$^{6}$,
C.C.~Miao$^{3,2}$,
J.~Niu$^{3,2}$, 
Z. C.~Pan$^{2}$,
L.~Qian$^{2}$,
M.Y.~Xue$^{2}$,
\newauthor
X.Y. Xie$^{6}$,
S.~P.You$^{6}$,
X.~H.~Yu$^{6}$,
M.~Yuan$^{3,2}$,
Y.L.~Yue$^{2}$,
Y.~Zhu$^{2}$ for the
\newauthor
CRAFTS collaboration\thanks{\url{https://crafts.bao.ac.cn/craftsTeamMem/}}\\
$^{1}$Max-Planck-Institut f{\"u}r Radioastronomie, Auf dem H{\"u}gel 69, D-53121 Bonn, Germany\\
$^{2}$National Astronomical Observatories, Chinese Academy of Sciences, Beijing 100101, China\\
$^{3}$CSIRO Astronomy and Space Science, PO Box 76, Epping, NSW 1710, Australia\\
$^{4}$Centre for Astrophysics and Supercomputing, Swinburne University of Technology, Mail H39, PO Box 218, VIC 3122, Australia.\\
$^{5}$ARC Center of Excellence for Gravitational Wave Discovery (OzGrav), Swinburne University of Technology, Mail H11, PO Box 218,\\ 
VIC 3122, Australia.\\
$^6$ Guizhou Normal University, Guiyang 550001, China\\ 
$^7$ University of Chinese Academy of Sciences, Beijing 100049, China\\ 
$^8$ NAOC-UKZN Computational Astrophysics Centre, University of KwaZulu-Natal, Durban 4000, South Africa\\
}

\date{Accepted XXX. Received YYY; in original form ZZZ}

\pubyear{2015}

\begin{document}
\raggedbottom
\label{firstpage}
\pagerange{\pageref{firstpage}--\pageref{lastpage}}
\maketitle

\begin{abstract}
We report the follow-up of 10 pulsars discovered by the Five-hundred-meter Aperture Spherical radio-Telescope (FAST) during its commissioning. The pulsars were discovered at a frequency of 500-MHz using the ultra-wide-band (UWB) receiver in drift-scan mode, as part of the Commensal Radio Astronomy FAST Survey (CRAFTS). We carried out the timing campaign with the 100-m Effelsberg radio-telescope at L-band around  1.36\,GHz. Along with 11 FAST pulsars previously reported, FAST seems to be uncovering a population of older pulsars, bordering and/or even across the pulsar death-lines. We report here two sources with notable characteristics. PSR~J1951$+$4724 is a young and energetic pulsar with nearly 100\% of linearly polarized flux density and visible up to an observing frequency of 8\,GHz. PSR~J2338+4818, a mildly recycled pulsar in a 95.2-d orbit with a Carbon-Oxygen white dwarf (WD) companion of $\gtrsim 1\rm{M}_{\odot}$, based on estimates from the mass function. This system is the widest WD binary with the most massive companion known to-date.  Conspicuous discrepancy was found between estimations based on NE2001 and YMW16 electron density models, which can be attributed to under-representation of pulsars in the sky region between Galactic longitudes $\ang{70}<l<\ang{100}$. This work represents one of the early CRAFTS results, which start to show potential to substantially enrich the pulsar sample and refine the Galactic electron density model.

\end{abstract}

\begin{keywords}
survey -- stars: neutron -- pulsar: general
\end{keywords}



\section{Introduction}
Pulsars are a type of rotating neutron star (NS) that emit beams of electromagnetic radiation along their magnetic axis. Pulsars are used to study stellar evolution \citep{Tauris2017}, to place limits on the equation of state for ultra-dense matter \citep{Lattimer2001,Ozel2016}, to map the free electron distribution of our Galaxy \citep{NE2001, YMW16}, and are remarkable natural laboratories for testing theories of gravity in the strong-field regime \citep{Kramer2006,Wex2014}.  Furthermore, besides being a source of emission of gravitational waves (GWs), such as the ones detected by LIGO/Virgo \citep{LIGO2017}, pulsars can also be used as tools to detect low-frequency GWs originating from supermassive-black-holes binaries. This is achieved through pulsar timing arrays \citep{Hobbs2010,Kramer2013, Janet2009}, in which the most rotationally-stable pulsars -- the precision of which approaches that of atomic clocks on long timescales -- are used as a window into the extremely low-frequency (${\sim}10^{-9}$ Hz) GWs, complementing the frequency ranges explored by LIGO/Virgo, LISA, and CMB experiments. To advance our understanding of the aforementioned fundamental physics, finding more pulsars is a priority.

A major full-visible-sky ($\sim$57\% of the whole sky) blind multi-purpose survey, namely the Commensal Radio Astronomy FAST survey (CRAFTS: \citealt{li18}) has started using the largest single-dish radio telescope, the Five-hundred-meter Aperture Spherical radio Telescope (FAST: \citealt{Nan2011,Qian2020}). Pulsar searching is a key component of CRAFTS, along with HI imaging, HI galaxies, and transients. Large-scale commensal survey of HI and pulsars has never been done before CRAFTS, mainly due to HI-imaging requiring  a regular injection of an electronic calibration signal to calibrate out short-time-scale gain variation.  Such a calibration will pollute the power spectrum with its harmonics and render new pulsars (periodical signals) hard to detect. We devised a high-cadence calibration injection technique by matching the injecting cadence with the sampling rate, which enables CRAFTS to simultaneously record HI and pulsar data streams. Pilot CRAFTS scans started in the middle of 2017 during the commissioning phase of FAST, with an un-cooled receiver, namely the Ultra-Wide-Band receiver (UWB: 270 MHz -- 1.62 GHz), which was proposed by \citet{Li2016} and \citet{li13}. In 2018, CRAFTS switched to using the L-band Array of 19-beams (FLAN: \citealt{zhang19-crafts}). More than 120 new pulsars have been confirmed by CRAFTS\footnote{\label{fastcat}\url{http://crafts.bao.ac.cn/}} to date.

The 100-m Effelsberg (EFF) radio telescope has been key in confirming and following-up the CRAFTS discoveries. Although much more sensitive, FAST needs to operate more than 1000 actuators to accomplish re-pointing, taking up to 10 minutes. The Effelsberg multi-beam system provides more time-effective and prompt sky coverage to help refine the location of the pulsar candidates, the uncertainty of which can be substantially bigger than the beam size during drift scans. 

\begin{figure*}
\begin{center}
\includegraphics[height=0.5\linewidth, angle=0]{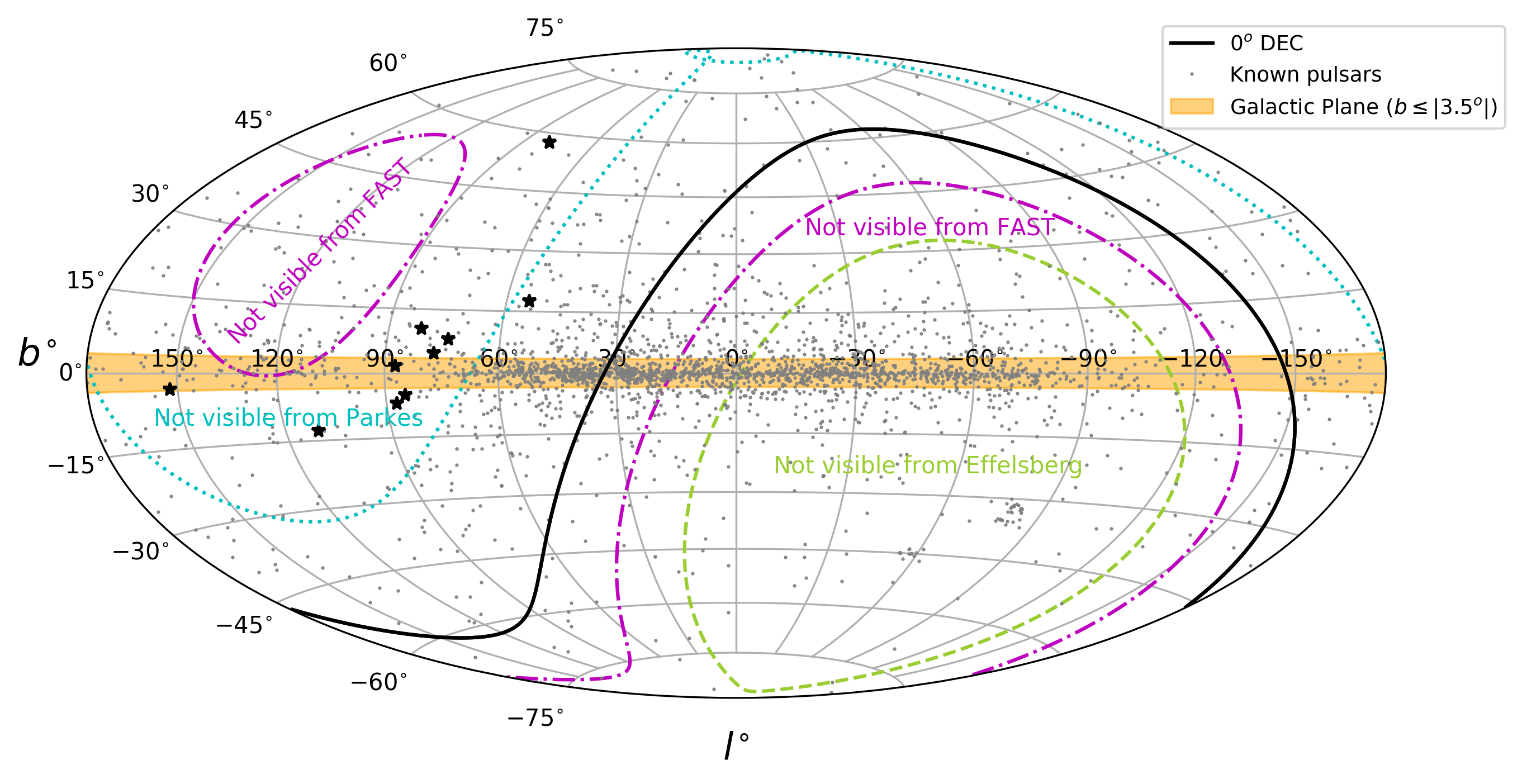} 
\end{center}
\caption{Sky coverage in Galactic coordinates of the 500-m FAST (magenta), the 100-m Effelsberg (green) and the 64-m Parkes (cyan) radio telescopes. The \textit{black-stars} show the position of the FAST/EFF pulsars described in Tables~\ref{tab:timing1},\ref{tab:timing2},\ref{tab:timing3} and \ref{tab:timing4}. The yellow shaded region shows the Galactic Plane ($\left|\text{b}\right|\leq\ang{3.5}$), the black curve shows the line of zero declination and the \textit{grey dots} show the known pulsars reported in the ATNF catalogue.}\label{fig: skymap} 
\end{figure*}

In this paper, we report the follow-up campaign with Effelsberg of 10 pulsars discovered by FAST as part of the CRAFTS pilot survey.
Along with previously published Parkes (PKS) radio telescope follow up paper \citep{Cameron2020}, these works represent 
the initiation of systematic efforts to build a new full sky pulsar sample (Figure~\ref{fig: skymap}). 
We describe the technical characteristics of the survey and the strategy employed for the follow-up observations in Section~\ref{sec:obs}, the timing methods and how to constrain the pulsar's geometry based on pulse profiles in Section~\ref{sec:methods}, the timing solutions and the individual features in Section~\ref{sec:results}.  Section~\ref{sec:discussion} presents the combined FAST-UWB new pulsar sample timed by EFF and PKS. We discuss the population of pulsars traced and why they have been missed by previous surveys. Our concluding remarks are given in Section~\ref{sec:conclusions}.
\section{Observations}\label{sec:obs}

\begin{figure}
\begin{center}
\includegraphics[height=0.55\linewidth, angle=0]{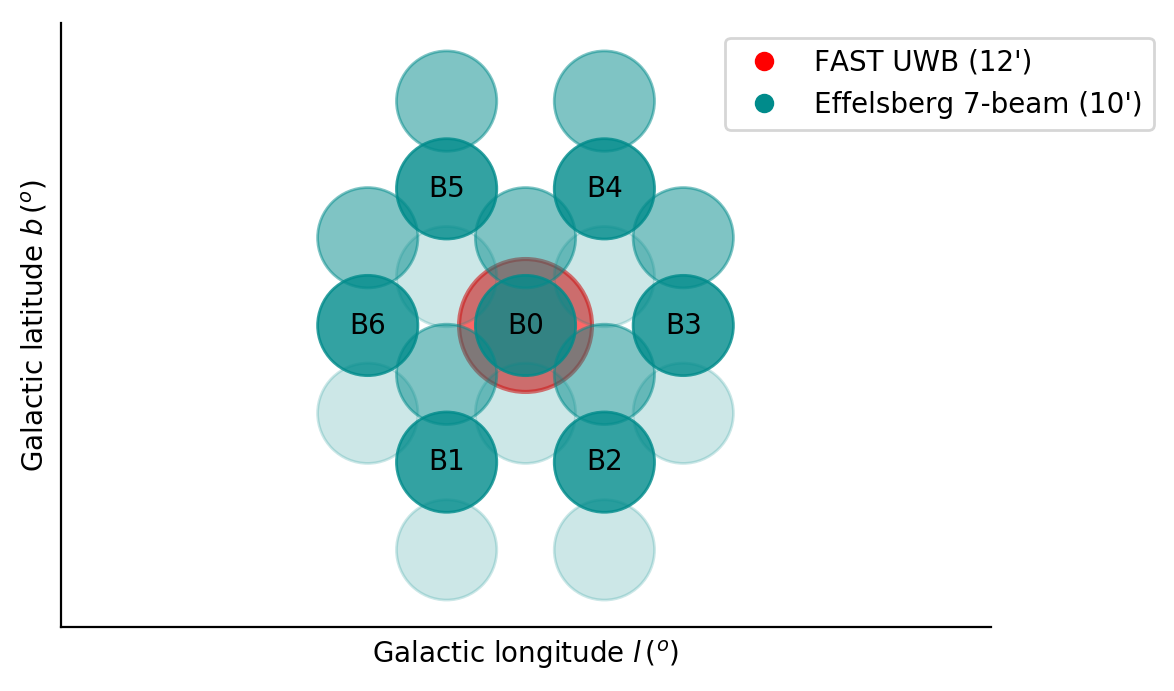} 
\end{center}
\caption{Effelsberg's 7-beam receiver setup and FWHM at 1.36 GHz and comparative size with FAST's UWB receiver FWHM at 0.5 GHz. The survey grid with the 7-beam receiver is constructed by applying a rotation of $\ang{30}$ to the start positioning angle to each of the 3 pointing needed to fully cover the region. An offset of $\pm\ang{0.14}$ is applied to the upper and lower pointing respectively.}\label{fig:7beam_setup} 
\end{figure}

\subsection{CRAFTS pulsar survey}
The CRAFTS pulsar survey searches the sky in drift-scan mode from declination $-14^\circ$ to $+66^\circ$ (see Figure~\ref{fig: skymap}). During the commissioning phase, it made use of a single-beam UWB receiver recording radio frequencies from 270\,MHz to 1.62\,GHz \citep{Li2016}. The UWB was an uncooled, testing instrument developed only for commissioning purposes: it allowed FAST to work on the data recording, positioning calibration, testing the timing precision as well as to work on the data flow and processing scripts. Various data taking parameters were tried with the FAST-UWB. We finally settled down on 100 $\mu$S sampling time and 0.5-MHz frequency channel width. The representative system temperature at 500\,MHz was about 70\,K. 
With the UWB receiver, most of the candidates were found in the 500\,MHz band (270-730\,MHz). In this band, a point radio source passes through the beam roughly in 50 seconds and the main beam has a full width at half maximum (FWHM) of 12 arcminutes (at 500 MHz, see Figure~\ref{fig:7beam_setup}). The UWB has led to the discovery of over 80 new pulsars, while a systematic reprocessing is on-going.

The short effective integration time of drift scans coupled with the immense raw sensitivity of FAST facilitate finding pulsars in relativistic binaries. The longer the integration time required to achieve a certain flux threshold, the larger number of  acceleration trials needed to account for the change during the observation in the apparent spin period of pulsars in binary systems \citep{Johnston1991,Andersen2018}. \citet{Smits2009} estimated that $\sim$4000 new pulsars would be accessible to FAST in the Galactic plane and \citet{liu18} estimated about 1000 in full sky for FAST drift scans. Such a high number of pulsars is hard (if not impossible) to follow-up with a single telescope. Additional telescope facilities play an important role in helping with the follow-up observations that lead to phase-connected timing solutions, the key to extracting the scientific potential of new pulsars. 

\subsection{Effelsberg follow-up}\label{subsec: Eff follow-up}    
Our Effelsberg follow-up observations are split into \textit{searching mode} and \textit{timing mode}. For both, we made use of the 7-beam receiver at a central observing frequency of 1.36\,GHz. For the search observations, we have used the Pulsar Fast Fourier Transform Spectrometer \citep[PFFTS,][]{Barr2013}.  The PFFTS records data with a time resolution of 54\,$\mu$s over a 300\,MHz bandwidth split into 512 frequency channels. However, the PFFTS is not synchronized with a maser clock. To compensate, we recorded simultaneously the data with the use of the high precision pulsar timing backend PSRIX \citep{Lazarus2016}. PSRIX has a bandwidth of 250 MHz divided into 256 frequency channels and records data in two polarizations. For the search mode, we made use of each of the beams of the 7-beam feed array, while for the timing observations only the central beam was employed.

In order to successfully confirm a candidate several factors had to be taken into account:
\begin{enumerate}[]
    \item \textbf{\textit{Grids}}. Positional uncertainties were expected due to the drift scan mode and the complexity of FAST's positioning system \citep{Nan2011}. Additionally, as is shown in Figure~\ref{fig:7beam_setup}, there is a difference in the FWHM, of 10$'$ and  12$'$ for Effelsberg and FAST beam size, respectively. To account for such offsets we have used the 7-beam receiver to construct a 3-pointing grid as displayed in Figure~\ref{fig:7beam_setup}, which covers a region of $\ang{0.48} \times \ang{0.66}$ around a given candidate in full beam-sampling. We rotated the 3-pointing grid along the declination axis of each candidate to maximize for declination offsets, which were the main source of error in the position. We refer to this beam set up as the \textit{search-grid}.
    
    \item \textbf{\textit{Telescope gain.}} Effelsberg's gain of 1.6 K/Jy (see \citealt{Cruces2020} for Effelsberg's 7-beam receiver sensitivity) versus the gain of 10 K/Jy of FAST's UWB receiver, implied that in order to achieve the same sensitivity an integration $\sim$40 times longer had to be considered. However, some candidates appeared to be significantly brighter than expected, mainly due to better pointing, and
    allowed for a shorter integration.
    
    \item \textbf{\textit{Spectral index}}. All the candidates reported in this paper came  from the lower part of the band of FAST's UWB receiver due to strong radio frequency interference (RFI) affecting frequencies above 800 MHz. To account for the decrease of the pulsar flux density when observed at higher frequencies -- such as L-band -- we assumed a spectral index of $-$1.6 \citep{Bates2013}.
    
    \item \textbf{\textit{Parameter space.}} Our parameter space is significantly reduced when compared with blind surveys due to the prior information about the pulsar, such as position, dispersion measure (DM), and spin period. However, the long integration times for Effelsberg meant that the data needed to be searched thoroughly in the acceleration space because of the change in the spin period due to the frequency shift caused by the orbital motion of pulsars in a binary system. To that end, we made use of the pulsar search algorithms implemented for the High Time Resolution Universe survey (see \citealt{Cherry2014, Barr2013} for survey description and search methods) based on: {\sc SIGPROC}'s\footnote{\url{http://www.pulsarastronomy.net/wiki/Software/Sigproc}} \textit{acceleration search}, a Fourier domain periodicity search where the time series is re-sampled at different constant acceleration trials; \texttt{RIPTIDE}\footnote{\url{https://github.com/v-morello/riptide}} implementation of a \textit{Fast Folding Algorithm}, a direct folding of the data at a range of trial periods \citep{Morello2020}; and {\sc PRESTO}'s \texttt{singlepulse.py}\footnote{\url{https://github.com/scottransom/presto}} \citep{Ransom2011} to search for single pulses in the time series. 
\end{enumerate}
These considerations meant that for a given FAST candidate, the observations carried out by Effelsberg ranged between 0.5 to 2 hours per pointing. Due to the long integrations needed, we were restricted to the binary systems that we could detect through \textit{acceleration search} according to the 10\% rule \citep{Ransom2002}. This means that given a 30 minutes observation needed to reach a signal-to-noise ratio (SNR) to allow for a confirmation, we were not sensitive to detect binaries with orbital periods shorter than $\sim$5 hours. We observed a total of $\sim$40 pulsar candidates, out of which 10 were confirmed and subsequently followed-up. The data of the non-confirmed candidates was further explored with \textit{jerk search}, a Fourier domain method implemented as part of the \textsc{PRESTO} search package, that accounts for a linearly changing acceleration within the observation. Non-confirmations can be attributed to the decrease of the flux at our observing frequencies, errors in coordinates introduced by the drift-scan survey, scintillation effects, and ultimately the difference in sensitivity between FAST and Effelsberg.

After a candidate was confirmed, a second search mode observation was performed soon after to fine-tune its position through a closely-packed-grid setup made only with the use of the central beam of the 7-beam receiver. This refined position allowed us to locate the pulsar somewhere closer to the center of our beam, enhancing the SNR.

We emphasize that all the considerations described above apply exclusively to the early commissioning phase with the UWB receiver. Different considerations were made after FAST was upgraded in May 2018 with the 19-beam receiver \citep{Jiang2020}. For that setup, a source passes through the 3$'$ beam in $\sim$20 seconds, and a gain of 18 K/Jy at L-band was reached. A report on the Effelsberg follow-up of the new discoveries from the FAST pulsar search with the 19-beam is in preparation. We refer hereon to the pulsars discovered by FAST and followed by Effelsberg as the FAST/EFF sample, and to the FAST pulsars followed by Parkes as the FAST/PKS sample.

\section{Method}\label{sec:methods}
\subsection{Phase connected timing solutions}\label{subsec: timing}
Despite the PFFTS's data being unreliable for timing due to the lack of precise time stamps, its filterbank format allows the observation to be easily searched for the pulsar and to measure its spin period at a given epoch in an incoherent de-dispersion mode. We thus combined the information from the initial confirmation with the subsequent observations to refine its position (see Section~\ref{subsec: Eff follow-up}) and were able to create a folding ephemeris with spin period and position accurate enough to record coherently de-dispersed and folded data (referred to as timing data hereafter) from an isolated slow pulsar.

As the folding ephemerides were robust on days-to-weeks scale, we planned multiple observations within that time-span. Times of arrival (TOAs) were created with the use of {\sc psrchive}\footnote{\url{http://psrchive.sourceforge.net/}} \citep{vanStraten2012}. With the use of {\sc tempo}\footnote{\url{http://tempo.sourceforge.net/}} and {\sc tempo2}\footnote{\url{https://bitbucket.org/psrsoft/tempo2}} \citep{Hobbs2006} the residuals of the timing model were minimised by fitting the spin frequency (F0) and its derivative (F1) if they were $3\sigma$ or more significant. As the observation continued for several months we fitted as well for the position of the source. If continuous observations were not phase connected we used the ``jumps" and ``phase $+n$" method in {\sc tempo} -- where $n$ is an integer number that accounts for the number of rotations missed --  to find whether any value of $n$ produces an unambiguous connection between TOAs. If this is not the case, we resorted to the use of {\sc dracula}\footnote{\url{https://github.com/pfreire163/Dracula}} \citep{Freire2018} to systematically bridge multiple gaps between TOAs.

For the binary pulsar in our sample, PSR~J2338$+$4818, in addition to the parameters mentioned above, it was necessary to establish an orbital solution before the start of the coherent timing data recording. This solution has five Keplerian parameters: the orbital period (PB), the epoch of periastron (T0), the projected semi-major axis (A1), the longitude of the periastron (OM), and the eccentricity (ECC). These were first estimated by measuring the change of the spin period at several epochs. Once this solution
is established, then using the timing data we can determine the phase connection. To do this we needed to use the {\sc dracula} routine, because of the sparsity of detections for this pulsar.

\subsection{Polarization calibration}\label{subsec:pol_cal}
We have carried out noise diode observations for several timing epochs to perform polarimetric calibration. The calibrator consists of an injected 100\% linearly polarized diode signal at 1 Hz and $\ang{45}$ to the feeds, at a sky region 0.5$^o$ offset from the source. Each polarimetric calibration  consisted of 90--120 seconds prior to each pulsar observation. The calibration was applied through the use of {\sc psrchive}'s \texttt{pac} routine by creating a database of calibrators for each pulsar at each epoch. The outcome of the procedure is observed in the pulse profiles shown in Figure~\ref{fig:pulse_profiles}. To this end, we have combined several epochs for each pulsar to obtain a profile of high signal-to-noise ratio. The integration times are listed in Table~\ref{tab:pol_prof}. The polarization properties reported in this work have adopted the instrumental convention described in \citet{vanStraaten2010}.

After correcting for Faraday rotation (see below), we measured the fraction of linear ($L/I$) and circular polarization relative to the total intensity from the on-pulse region. We list these values in Table~\ref{tab:pol_prof} as well. For the circular polarization, we determined the relative $V/I$ and absolute $\left|V\right|/I$ values.

\subsection{Rotation measure (RM) determination}\label{subsec:rm}
After applying the polarization calibration to the data, we corrected the effect of Faraday rotation on the linear polarization. Assuming a stable RM, we report the values obtained from the combined set of observations for each pulsar (see Table~\ref{tab:pol_prof}).

We used two approaches to measure the RM value. The first is \texttt{rmfit} from {\sc psrchive}, which is based on an optimization of the linear polarization fraction solely by using a range of RM trials. For comparison we also used \texttt{RMcalc.py} code implemented by \citet{Porayko2019}. This approach is based on the RM synthesis method described in \citet{Brentjens2005}, and reports estimates from the Bayesian Generalised Lomb-Scargle Periodogram (BGLSP) technique applied to the RM synthesis method when the RM is unambiguously determined. The RMs determined trough both methods are presented in Table~\ref{tab:pol_prof}.

We have applied the corresponding Faraday rotation corrections to each pulsar profile only when linear polarization was measured and the RM value was above $1\sigma$ significant and in agreement for both RM determinations.

\subsection{Rotating vector model}\label{subsec:rvm}
We derived the geometry of the pulsars in the FAST/EFF sample whenever the results of the polarization observations led to a well-defined position angle (PA). We used the rotating vector model (RVM) described by \cite{Radhakrishnan69} to relate the PA to the projection of the magnetic inclination angle $\alpha$ and the impact parameter $\beta$ through \citep{Handbook2012}:
\begin{equation}
\rm{tan}(\psi_o-\psi)=\dfrac{\rm{sin}(\alpha) \,sin(\phi_o-\phi)}{\rm{sin}(\zeta)\rm{\,cos}(\alpha)-\rm{cos}(\zeta)\rm{\,sin}(\alpha)\rm{\,cos}(\phi_o-\phi)}\label{eq:rvm} \> ,
\end{equation}
 where $\phi$ is the rotational phase, $\zeta=\alpha+\beta$ (viewing angle) and $\phi_o$  and $\psi_o$ correspond to the rotational phase and position angle corresponding to the fiducial plane respectively. To constrain the pulsar's viewing geometry we performed a Markov Chain Monte Carlo (MCMC) fit to the parameters in Equation~\ref{eq:rvm}.

\section{Results}\label{sec:results}
\begin{figure*}
\begin{center}
\includegraphics[width=0.7\linewidth, angle=0]{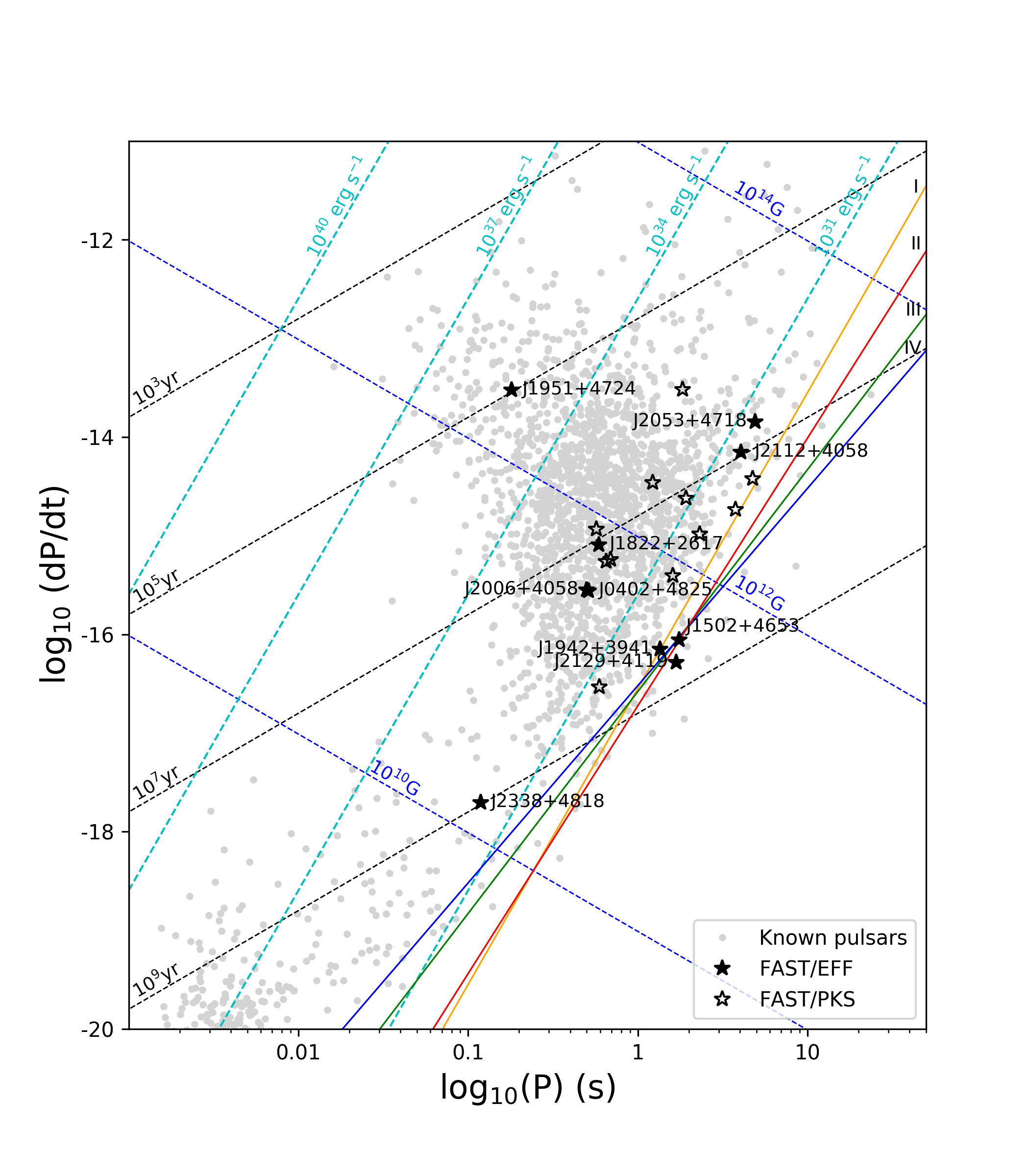} 
\end{center}
\caption{A $P-\dot{P}$ diagram. Known pulsars are shown with \textit{grey dots}, while the new pulsar discoveries reported in this work are plotted with \textit{filled-black stars}, and the pulsars reported by \citet{Cameron2020} with \textit{open-black stars}. Alongside are drawn lines of constant magnetic field strength (\textit{dark-blue dashed lines}), lines of constant spin-down age (\textit{black dashed lines}) and lines of constant rotational energy loss (\textit{cyan dashed lines}) as derived from the rotating dipole model. The death lines shown correspond to \citet{Bhattacharya1992} polar gap model (\textit{orange-line}; model I),  \citet{Chen1993} model for a decreased polar cap area (\textit{red-line}; model II), and \citet{Zhang2000} models for curvature radiation from the vacuum gap model (\textit{green-line}; model III) and from the space-charged-limited flow (\textit{blue-line}; model IV).}\label{fig:p-pdot} 
\end{figure*}

The timing solutions reported in Tables~\ref{tab:timing1}, \ref{tab:timing2}, \ref{tab:timing3} and \ref{tab:timing4} are the result of (at least) one year of timing follow-up. It is worth noting that in order to properly estimate uncertainties, we have scaled the TOA uncertainties by the so-called EFAC such that the residuals have a reduced $\chi^2\sim 1$. 

\begin{figure*}
\begin{center}
\includegraphics[trim={0 7cm 0 0},width=1\linewidth, angle=0]{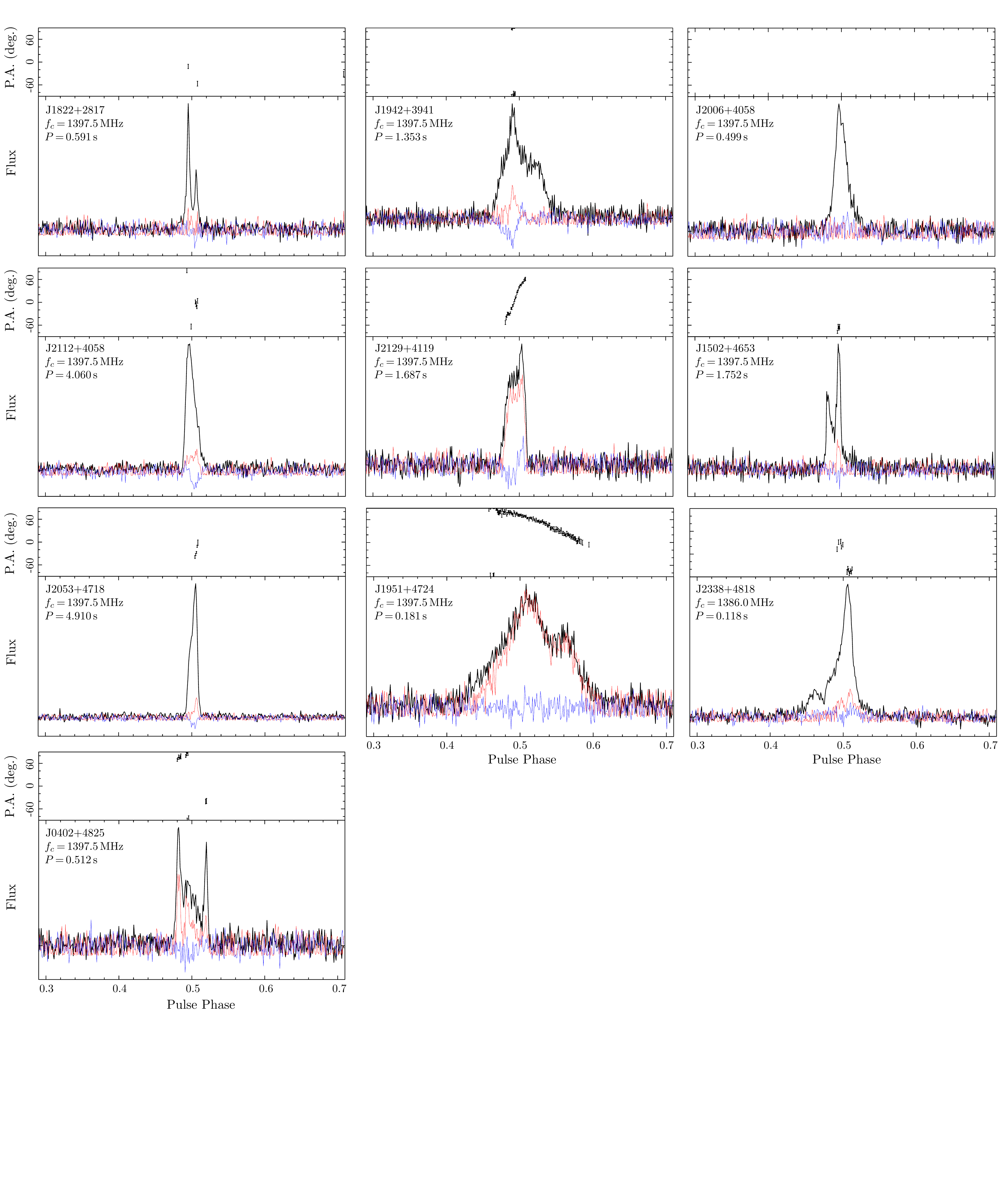} 
\end{center}
\caption{Calibrated average pulse profiles of the FAST/EFF pulsars. The polarization profiles were taken at L-band and have been corrected for Faraday rotation whenever RM was measured (see Table~\ref{tab:pol_prof}). The total intensity is shown in black, while linearly-polarized flux and circular-polarized flux are shown in red and blue respectively. The integration times used are listed in Table~\ref{tab:pol_prof}. Above each profile is shown the position angle as a function of the pulse phase.}\label{fig:pulse_profiles} 
\end{figure*}

Figure~\ref{fig:pulse_profiles} display the integrated pulse profiles (in black) with the linearly polarized and circularly polarized flux density in red and blue, respectively. The pulse profiles are corrected for the RM values reported in Table~\ref{tab:pol_prof} as measured from the RM synthesis technique described in Section~\ref{subsec:rm}. To find the RM we explored for each pulsar a range between [-10000, 10000] $\text{rad}\,\text{m}^{-2}$ and afterwards zoomed into the region around the peak RM value to refine the search. For some sources a limit on its RM is provided as trials within such RM range led to not significant change in the linear polarization fraction. 
We measure the pulse width at 50\,\% and 10\,\% of the profile peak, $W_{50}$ and $W_{10}$ respectively. Above each pulse profile is displayed the observed position angle (PA). To constrain the geometry of the pulsars we fit the PAs with the standard RVM through a Monte-Carlo simulation. With the exception of PSRs J2112$+$4058, J2129$+$4119, J1951$+$4724, and J0402$+$4825, it was not possible to constrain the geometry either because of the lack of observed polarization or due to insufficiently well-defined PA curves.

Regarding the sky location of the FAST/EFF sources we see in Figure~\ref{fig: skymap} that out of the 10 pulsars, 2 are located in the Galactic plane, 7 pulsars in the middle (mid) Galactic latitude region, and one at high latitude. More detailed population analysis is carried in Section~\ref{sec:discussion}. We proceed hereon with the summaries of the individual pulsars, which have been sorted in ascending declination.
\subsection{PSR J1822\texorpdfstring{$+$}\,2617}\label{subsec:J1822+2717}
PSR J1822$+$2617 is a 0.591-s pulsar with a dispersion measure of 64.7 $\text{cm}^{-3}\,\text{pc}$. It was first confirmed by the Parkes follow-up project  \citep{Cameron2020}, however, due to its high declination, it was transferred soon after to the Effelsberg campaign. As seen in the $P-\dot{P}$ diagram in Figure~\ref{fig:p-pdot}, it lies in the middle of the normal radio pulsar population.

It has a sharp double-peaked pulse profile, with the left component being the dominant source of emission. With the use of the cross-correlation function of the pulse dynamic spectra \citep{Main2017} over the best observation, we estimate the diffractive scintillation timescale to be $\Delta\tau=783\pm$9\,s and the scintillation bandwidth to be $\Delta\nu=4.9\pm$0.6\,MHz. The expectation from the \citet[][hereon NE2001]{NE2001} free electron density model for the line-of-sight of PSR J1822$+$2117 are $\Delta\tau=370\pm70$\,s and $\Delta\nu=1\pm$1\,MHz.

Regarding its RM, we find 125 $\text{rad}\,\text{m}^{-2}$ through \texttt{rmfit} and 67 $\text{rad}\,\text{m}^{-2}$ with RM synthesis. When applying the corresponding Faraday rotation correction, the use of one value over the other leads to no significant difference in the linear polarization fraction.

\subsection{PSR J1942\texorpdfstring{$+$}\,3941}\label{subsec:J1942+3941}
PSR J1942$+$3941 is a 1.353-s pulsar with the widest pulse profile from the FAST/EFF sample given its $\text{W}_{10}$ of 113.63\,ms. With the 8500\,s of integration time used to obtain the pulsar profile, a small fraction of linear and circular polarization of 7\% and 9\% respectively was found. Interestingly, the RM through RM synthesis was found to be consistent with 0 $\text{rad}\,\text{m}^{-2}$ within $3\sigma$, while \texttt{rmfit} constrained to be $<|100|$ $\text{rad}\,\text{m}^{-2}$. However, correcting the effect of Faraday rotation with the RM values estimated by both methods led to no significant differences in the linear polarization fraction. Observations at lower frequencies and wide bandwidths, for instance with the Low-Frequency Array (LOFAR), could provide better constraints.

The distance inferred based on the pulsar's DM of 104.5 $\text{cm}^{-3}\,\text{pc}$ is 5.4 kpc and 8.5 kpc for the NE2001 and \citet[][hereon YMW16]{YMW16} models, respectively. Their disagreement is further discussed in Section~\ref{subsec:ed_models}.

\subsection{PSR J2006\texorpdfstring{$+$}\,4058}\label{subsec:J2006+4058}
PSR J2006$+$4058 is a 0.499-s pulsar with a DM of 259.5 $\text{cm}^{-3}\,\text{pc}$, and as seen in Figure \ref{fig:p-pdot}, is yet another example of a radio powered pulsar in the middle of the normal pulsar zone. Despite 14980-s of integration time, it was not possible to measure polarization. We split the observing bandwidth of 250 MHz into 10 sub-bands of 25 MHz each and explored whether the tail in the pulse profile was a consequence of scattering. We did not observe any significant frequency-dependent broadening of the pulse profile.

Interestingly, despite its low Galactic latitude of $b=4.73^o$, the NE2001 electron density model fails to estimate its distance and instead places a lower limit of $d$>50 kpc. If this was true, the pulsar would be located outside of the Galaxy. Contrary, the YMW16 model provides better constraints on the electron density along that line-of-sight and predicts a DM distance of 12.5 kpc.

\subsection{PSR J2112\texorpdfstring{$+$}\,4058}\label{subsec:J2112+4058}
With a rotational period of 4.060\,s, PSR J2112$+$4058 is the second slowest pulsar of the sample. Its pulsar profile displays a weak scattering tail which can be attributed to its DM of 129 $\text{cm}^{-3}\,\text{pc}$. We measured 15\% and 7\% of linear and circular polarization respectively.  Although there are few points in the PA swing, its inflection point ($\phi_o$) is well defined and hence we are able to put constraints on the geometry of PSR J2112$+$4058.  Through the RVM method described in Section~\ref{subsec:rvm}, we estimated the magnetic inclination angle to be $\alpha=\ang{94}^{+\ang{40}}_{-\ang{42}}$ ($1\sigma$ uncertainties). Given the large uncertainties for $\alpha$ (and $\zeta$), we were unable to constrain $\beta$. 

\subsection{PSR J2129\texorpdfstring{$+$}\,4119}\label{subsec:J2129+4119}
PSR J2129$+$4119  has a spin period of 1.687\,s and a low DM of 31 $\text{cm}^{-3}\,\text{pc}$. The pulsar scintillates with a diffractive scintillation timescale $\Delta\tau=50\pm7$\,s over a bandwidth of $\Delta\nu=6\pm$0.7\,MHz, as inferred from the cross-correlation function of the pulse dynamic spectra. PSR J2129$+$4119 is one of the nearest pulsars from the sample, with the NE2001 model predicting a DM distance of 2.3 kpc, and the YMW16 estimating it to be 1.9 kpc.

\begin{figure*}
\begin{center}
\includegraphics[trim={2cm 0cm 2cm 0cm},width=\linewidth, angle=0]{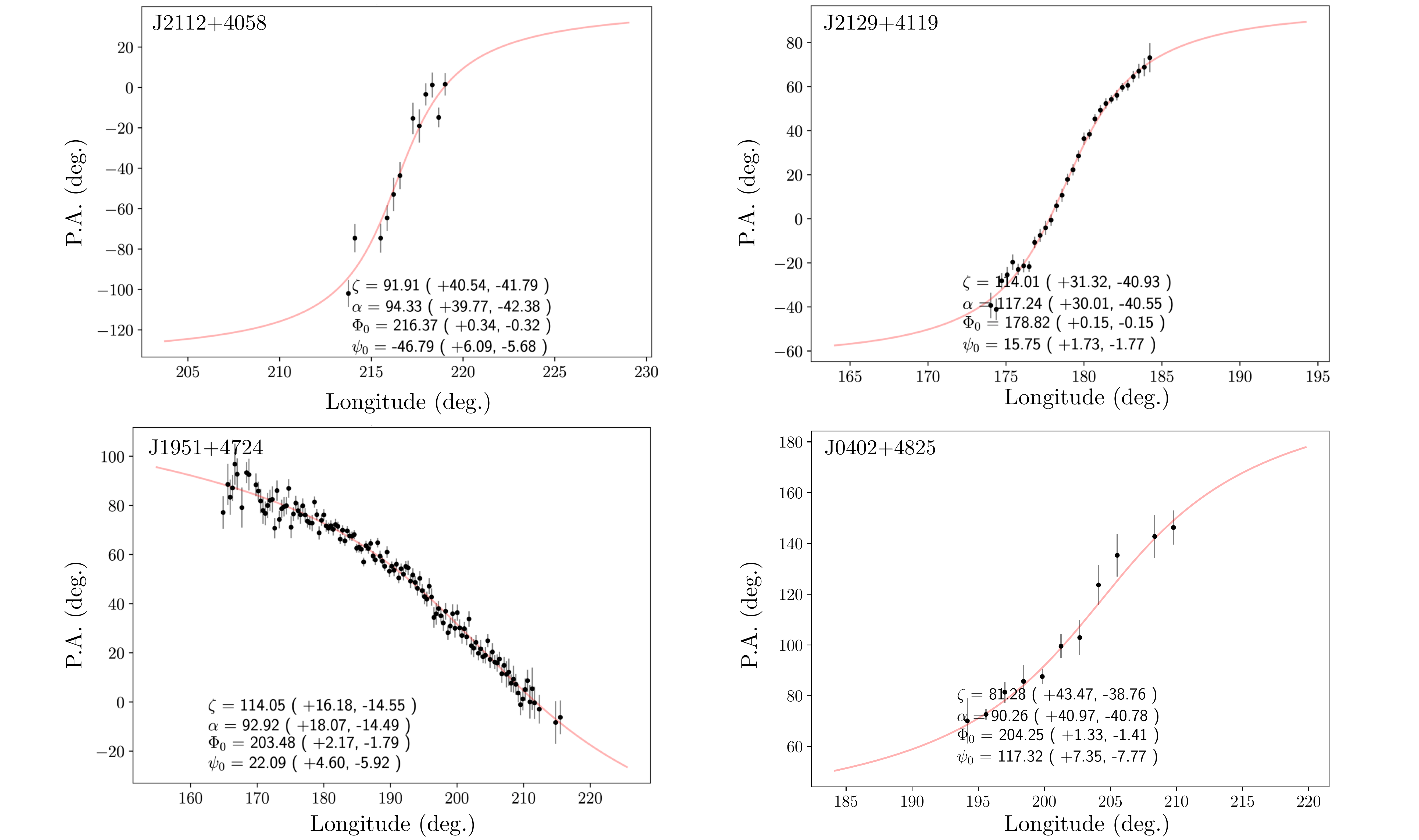} 
\end{center}
\caption{Viewing geometry of PSRs J2112$+$4058 (\textit{top-left}), J2129$+$4119 (\textit{top-right}), J1951$+$4724 (\textit{bottom-left}) and J0402$+$4825 (\textit{bottom-right}). In each plot, the red-curve corresponds to the least-squares fit to the polarization angle displayed in Figure\,\ref{fig:pulse_profiles}. $\psi_0$ is the polarization position angle, $\phi_0$ the rotational phase, $\alpha$ the magnetic inclination angle and $\zeta$ the viewing angle. The mean value of the RVM posterior distribution for each parameters is displayed with its 1$\sigma$ uncertainties.}\label{fig:RVMfit} 
\end{figure*}

From the pulse profile shown in Figure~\ref{fig:pulse_profiles}, we see that PSR J2129$+$4119 shows a high degree of linear polarization (>70\%). Its PA swing is remarkably well defined over its entire pulse phase and allows the geometry to be constrained.  As shown in Figure~\ref{fig:RVMfit}, we find $\alpha=\ang{117}^{+\ang{31}}_{-\ang{41}}$  and again, given the uncertainties for $\alpha$ and $\zeta$ we are not able to constrain $\beta$.

\subsection{PSR J1502\texorpdfstring{$+$}\,4653}\label{subsec:J1502+4653}
PSR J1502$+$4653 is a sharp double-peaked pulsar spinning at a period of 1.752 seconds. The right component of the pulse profile is dominant and is the only one showing traces of linear polarization.
Using an integration of 53970 seconds we measured a degree of linear polarization of 8\%.

With a DM of 26.6 $\text{cm}^{-3}\,\text{pc}$, the NE2001 model predicts a DM distance of 1.5 kpc while the YMW16 model puts a lower limit to the distance of 25 kpc. The disagreement between both estimations clearly shows the lack of constraints for pulsars at high Galactic latitudes due to the under-representation of sources in this area, which limits the mapping of the free electron distribution. This can be seen in Figure~\ref{fig: skymap}, where in addition to the FAST/EFF pulsars, the population of the known pulsars is shown.

\subsection{PSR J2053\texorpdfstring{$+$}\,4718}\label{subsec:J2053+4718}
With a period of 4.910\,s PSR J2053$+$4718 is the slowest pulsar of the FAST/EFF sample. The RM deduced from RM synthesis is $-785\pm1\,\text{rad}\,\text{m}^{-2}$, which is in agreement with the linear polarization optimization carried by \texttt{rmfit}. Furthermore, the Bayesian Generalized Lomb-Scargle Periodogram (BGLSP) method applied to the standard RM synthesis technique \citep{Porayko2019} gives $-786\pm14\,\text{rad}\,\text{m}^{-2}$. Considering that this pulsar is located close to the Galactic plane ($b=1.6174$) and that it is in the Cygnus area, where there are a several supernova remnants and excess gas, its DM of 331.3\,$\text{cm}^{-3}\,\text{pc}$ and the high RM is perhaps expected. 

Regarding the surrounding area in the Galactic plane (see Figure~\ref{fig: skymap}), we see that is not densely populated. This may be the reason why the NE2001 model fails to provide an estimation of its DM distance and instead puts an upper limit of roughly 50\,kpc. Again, this would place the pulsar outside of the Galaxy. On the other hand, the YMW16 model predicts the distance to be 8.9 kpc.

\subsection{PSR J1951\texorpdfstring{$+$}\,4724}
PSR J1951$+$4724  is the youngest and most energetic pulsar of the dataset. With a spin period of 0.181\,s and a spin period derivative of $3.01247(9)\times10^{-14}$, the pulsar is located at the top left of the $P-\dot{P}$ diagram. Because of its high spin-down luminosity of $1.5\times10^{35} \text{erg s}^{-1}$, we inspected the Fermi-LAT 10-Year Point Source Catalog\footnote{\url{https://heasarc.gsfc.nasa.gov/W3Browse/fermi/fermilpsc.html}} to search for a $\gamma$-ray counterpart in the 50 MeV to 1 TeV  energy range. Nonetheless, we found no counterpart within 1 degree.

From the pulse profile shown in Figure~\ref{fig:pulse_profiles} we see that PSR J1951$+$4724 has a degree of linear polarization of 90\%, the highest fraction of polarization among FAST/EFF pulsars. The well defined PA swing across the full pulse phase provides the tightest constraints of the geometry study. We find $\alpha=\ang{92}^{+\ang{19}}_{-\ang{14}}$ and $\zeta=\ang{114}^{+\ang{16}}_{-\ang{15}}$.

We observed the J1951$+$4724 with the C$+$ receiver at Effelsberg, covering  3 to 8 GHz continuously. PSR J1951$+$4724 is detected up to the top of the band. Further analysis of its spectral index will be the subject of future work. Although a high degree of linear polarization was detected in this band, no measurable RM was found. PSR J1951$+$4724 is relatively bright with a modest DM of 104.35 $\text{cm}^{-3}\,\text{pc}$, a potentially suitable target for  LOFAR, which should  better constrain its RM value.
\subsection{PSR J2338\texorpdfstring{$+$}\,4818}\label{subsec:C01}
PSR J2338$+$4818 is a 118.7-ms pulsar, part of a wide binary system, with an associated DM of 35 $\text{cm}^{-3}\,\text{pc}$. It scintillates and shows evidence of long-term nulling \citep{Backer1970}.
\subsubsection{Binary system}
We observed a small change in the apparent spin period within a couple of months. A Lomb-Scargle periodogram analysis based on the measured spin periods at several epochs gave a 95-day orbital period. With this orbital period plus the amplitude of the oscillation in the spin period, we were able to roughly estimate the additional binary parameters; this was then
used to determine the phase-coherent timing solution, using the methods described by \citet{Freire2018}. Two years of observations were necessary in order to obtain enough TOAs; this was due to the low detection rate described in Section~\ref{subsubsec: scintillation}. 

The timing solution includes a highly precise estimate of the spin period (about 0.1187\,s), and a spin period derivative of $1.98\times10^{-18}$.
The orbit has an eccentricity of 0.0018. The mass function of the system is 0.19263(4) and assuming a  pulsar mass of 1.4 M$_{\odot}$, then the minimum mass of the companion star is 1.049 M$_{\odot}$. Given that minimum companion mass, the orbital parameters and the inferred age of the pulsar of $\sim0.95$\,Gyr, the companion is most likely a carbon-oxygen white dwarf (CO-WD) \citep{Tauris2011,Tauris2012}.

\subsubsection{Scintillation}\label{subsubsec: scintillation}
According to the NE2001 model the estimated diffractive scintillation timescale is expected to be ${\sim}10$ minutes for an assumed 100 km/s source velocity. We investigated the timescale for the scintillation by analyzing its dynamic spectrum at 1.36 GHz. Based on two observations with high SNR, we estimated the scintillation timescale to be $\Delta\tau=660\pm70\,$s, over a bandwidth of $\Delta\nu=18\pm3\,$MHz. This is in agreement with the estimation from the NE2001 model. Interestingly, the source was detected only in 42\% of the observations. Usually, each observation lasted an hour and often two or more observations were carried-out per observing epoch, thus, diffractive scintillation alone is unlikely to explain its frequent disappearance.

\subsubsection{Long-term nulling}
\begin{figure}
\begin{center}
\includegraphics[trim={0.8cm 0 0.3cm 0},width=\linewidth, angle=0]{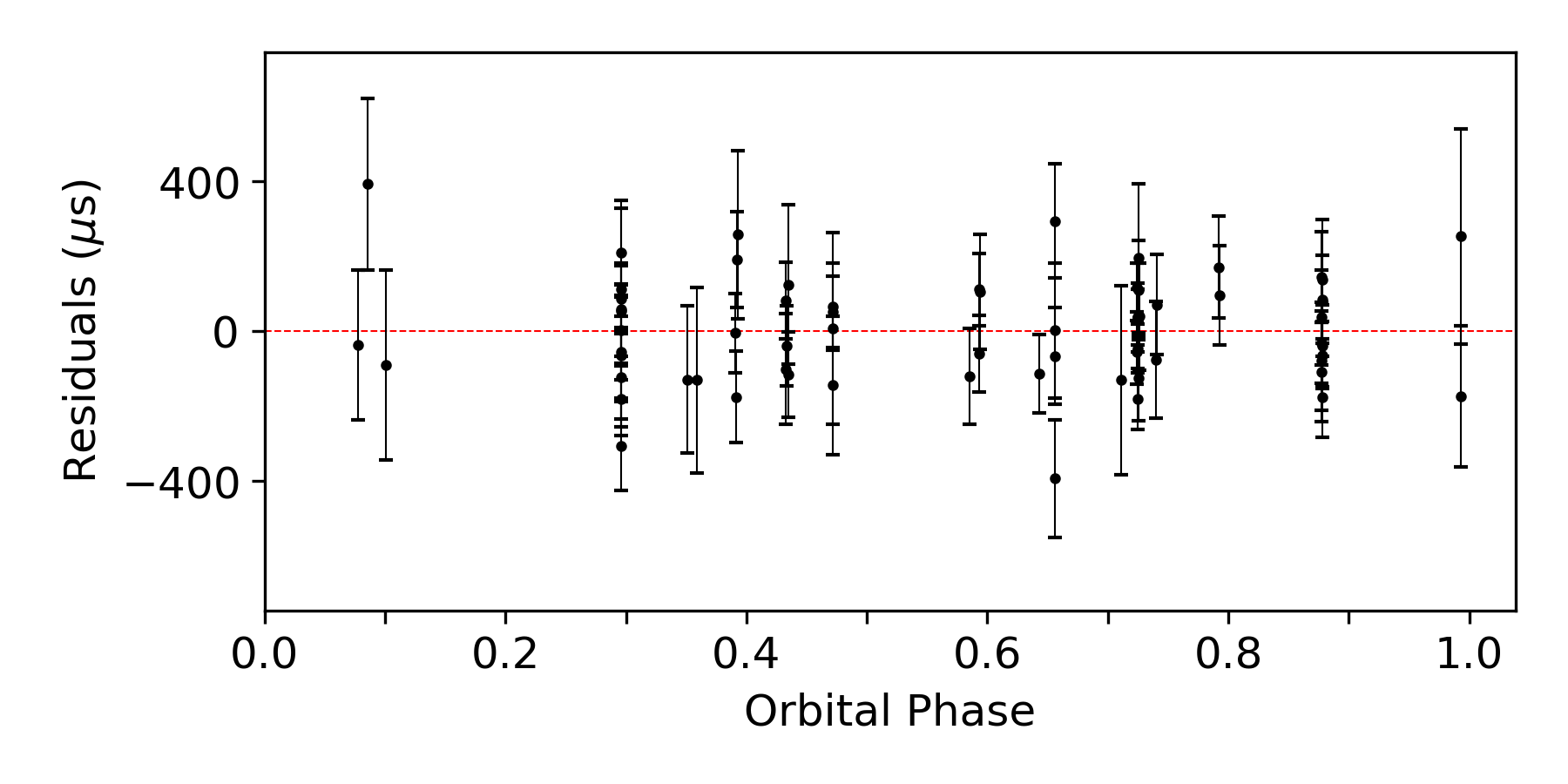} 
\end{center}
\caption{Timing residuals vs. binary orbit phase for PSR J2338$+$4818, based on the timing solution presented in Table\,\ref{tab:timing4}. The red-dashed line corresponds to the arrival time as predicted by the model. The error bars have been scaled based with an EFAC such that the reduced $\chi^2\sim1$.}\label{fig:J2338_res} 
\end{figure}

Over 30 search-mode filterbank files were recorded in parallel with timing observations of PSR~J2338$+$4818. Whenever the pulsar was not detected in the folded archives, we ran a targeted pulsar search over the filterbank files. In each case, we were also unable to detect the pulsar in the filterbank search data. This rules out a non-detection due to poorly modeled binary parameters. Furthermore, when two or more observations were recorded within a few hours, a non-detection in one observation was also followed by a non-detection in the others. We also recorded two continuous observations of roughly four hours each on different epochs. In the first epoch, the source was not detected, either in the full observation length or when processed in individual 30-minute chunks. During the second epoch, the source was detected during the full session. This suggests that the off-mode timescale is at least a few hours.
To exclude the possibility of a orbit-dependent detectability, we show in Figure\,\ref{fig:J2338_res} the timing residuals against the orbital phase. No noticeable trend is seen. We claim that the gap around phase 0.2 is due to the lack of data. Additional observations and further discussions on the detections of PSR J2338$+$4818 are part of a future paper.  

\subsection{PSR J0402\texorpdfstring{$+$}\,4825}
PSR J0402$+$4827 is a 0.512-s pulsar at a DM of 85.7 $\text{cm}^{-3}\,\text{pc}$. It displays an interesting pulse profile consisting of two sharp peaks and an intermediate component. We find a RM of $-111(70)\,\text{rad}\,\text{m}^{-2}$ and -89(9) $\,\text{rad}\,\text{m}^{-2}$  though \texttt{rmfit} and RM synthesis respectively. Both values are consistent and when applying the corresponding Faraday rotation correction, the use of one value over the other leads to no noticeable difference in the linear polarization fraction.

We see from the RM corrected pulse profile shown in Figure~\ref{fig:pulse_profiles} that each of the three components shows a fraction of linear polarization, with the main contribution to it being the leading component. From the PA we determine $\alpha=\ang{90}\pm\ang{41}$, but once more, we are not able to constrain $\beta$.

\begin{table*}
\begin{center}
\caption{Timing solutions part 1}\label{tab:timing1}
\begin{tabular}{llll}
\hline
Pulsar name & PSR J1822$+$2617  & PSR J1942$+$3941 & PSR J2006$+$4058 \\
\hline
\hline
\textit{Parameters} \\
Right ascension, $\alpha$ (J2000)\dotfill & 18:22:44.819(2)  & 19:42:22.05(3) & 20:06:39.098(3)   \\
Declination, $\delta$ (J2000)\dotfill & $+$26:17:26.83(4) & $+$39:41:41.4(7) & $+$40:58:53.48(3)  \\
Spin frequency, $F$ ($\text{s}^{-1}$)\dotfill & 1.690852663400(4)  & 0.73893939554(5)  &  2.001221096605(8)\\
Spin frequency derivative, $\dot{F}$ ($\text{s}^{-2}$)\dotfill &  $ -2.282(2) \times 10^{-15}$ & $-3.9(7) \times 10^{-17}$ & $-1.138(1) \times 10^{-15}$    \\
Spin period, $P$ (s)\dotfill & 0.591417585722(1) & 1.35329095461(9) & 0.499694912119(2)   \\
Spin period derivative, $\dot{P}$\dotfill & 7.983(6)$\times 10^{-16}$ & 7(1)$\times 10^{-17}$ & 2.843(4)$\times 10^{-16}$   \\
Dispersion measure, DM ($\text{cm}^{-3}\,\text{pc}$)\dotfill & 64.7(1) & 104.5(6) & 259.5(2) \\
\\
\hline
\textit{Fitting parameters} \\
First TOA (MJD)\dotfill &  58481.393 & 58462.761 & 58324.019   \\
Last TOA (MJD)\dotfill & 58895.203 & 58930.589  & 58839.727   \\
Timing epoch (MJD)\dotfill & 58688.00  & 58651.00 & 58582.00   \\
Number of TOAs\dotfill & 132 & 56 & 79  \\
Total integration time (s)\dotfill & 49400 & 58080 &  54436  \\
Weighted RMS residual ($\mu\text{s}$)\dotfill & 262.27 & 2881.102 & 469.06  \\
EFAC \dotfill & 1.304 & 1.361 & 1.021  \\
\\
\hline
\textit{Derived parameters} \\
Galactic longitude, $l$ ($^\circ$)\dotfill & 54.0268 & 73.6411 & 77.1185    \\
Galactic latitude, $b$ ($^\circ$)\dotfill & 17.5535 & 8.0654 & 4.7313   \\
DM distance, $d$ (kpc) \\
\multicolumn{1}{r}{NE2001} & 3.6 & 5.4 & 50 >   \\
\multicolumn{1}{r}{YMW16}  & 7.8 & 8.5 & 12.6   \\
Characteristic age, $\tau_\text{c}$ (Myr)\dotfill & 12.0 & 314.2 & 28.6   \\
Surface magnetic field, $B_\text{surf}$ ($10^{10}\,\text{G}$)\dotfill & 69.5 & 31.1 & 38.1   \\
Spin-down luminosity, $\dot{E}$ ($10^{30}\,\text{erg}\,\text{s}^{-1}$)\dotfill & 152.3 & 1.1 & 89.9   \\
\hline
\end{tabular}
\end{center}
\end{table*}
\begin{table*}
\begin{center}
\caption{Timing solutions part 2}\label{tab:timing2}
\begin{tabular}{llll}
\hline
Pulsar name & PSR J2112$+$4058  & PSR J2129$+$4119 & PSR J1502$+$4653  \\
\hline
\hline
\textit{Parameters} \\
Right ascension, $\alpha$ (J2000)\dotfill & 21:12:51.76(6) & 21:29:21.46(4) & 15:02:19.83(1)    \\
Declination, $\delta$ (J2000)\dotfill & $+$40:58:04(1) & $+$41:19:55(1) & $+$46:53:27.4(1)   \\
Spin frequency, $F$ ($\text{s}^{-1}$)\dotfill & 0.24625963556(3) & 0.59262128589(4) & 0.57061078387(1)    \\
Spin frequency derivative, $\dot{F}$ ($\text{s}^{-2}$)\dotfill & $-4.24(8) \times 10^{-16}$ &  $-3(2) \times 10^{-17}$ & $-5.6(3) \times 10^{-17}$   \\
Spin period, $P$ (s)\dotfill & 4.0607548114(5) & 1.68741829528(1) & 1.75250806373(3)   \\
Spin period derivative, $\dot{P}$\dotfill & 7.0(1)$\times 10^{-15}$ & 8(1)$\times 10^{-17}$ & 1.7(1)$\times 10^{-16}$   \\
Dispersion measure, DM ($\text{cm}^{-3}\,\text{pc}$)\dotfill & 129(8) & 32(1)  & 26.6(5)  \\
\\
\hline
\textit{Fitting parameters} \\
First TOA (MJD)\dotfill & 58522.51 & 58515.68 & 58431.88   \\
Last TOA (MJD)\dotfill & 58907.66 & 58839.63  & 58858.13    \\
Timing epoch (MJD)\dotfill & 58682.00 & 58695 & 58645.00   \\
Number of TOAs\dotfill & 33 & 60 & 100   \\
Total integration time (s)\dotfill & 44960 & 74900 & 144100  \\
Weighted RMS residual ($\mu\text{s}$)\dotfill & 1575.72 & 2270.86 & 775.61   \\
EFAC \dotfill & 1.32  & 1.235 & 0.677   \\
\\
\hline
\textit{Derived parameters} \\
Galactic longitude, $l$ ($^\circ$)\dotfill & 84.7385 & 87.1948  & 79.3056   \\
Galactic latitude, $b$ ($^\circ$)\dotfill & $-$5.1570 & $-$7.1093 & 57.6263   \\
DM distance, $d$ (kpc) \\
\multicolumn{1}{r}{NE2001} & 5.4 & 2.3 & 1.5   \\
\multicolumn{1}{r}{YMW16}  &5.2 & 1.9 & 25.0   \\
Characteristic age, $\tau_\text{c}$ (Myr)\dotfill & 9.2 & 342.8 & 167.0   \\
Surface magnetic field, $B_\text{surf}$ ($10^{10}\,\text{G}$)\dotfill & 539 & 37.2 & 55.2   \\
Spin-down luminosity, $\dot{E}$ ($10^{30}\,\text{erg}\,\text{s}^{-1}$)\dotfill & 4.1 & 0.65 & 1.2   \\
\hline
\end{tabular}
\end{center}
\end{table*}
\begin{table*}
\begin{center}
\caption{Timing solutions part 3}\label{tab:timing3}
\begin{tabular}{llll}
\hline
Pulsar name & PSR J2053$+$4718  &  PSR J1951$+$4724& PSR J0402$+$4825  \\
\hline
\hline
\textit{Parameters} \\
Right ascension, $\alpha$ (J2000)\dotfill & 20:53:45.49(4) & 19:51:07.45(3) & 04:02:40.633(9)   \\
Declination, $\delta$ (J2000)\dotfill & $+$47:18:55.3(4)  &   $+$47:24:35.1(2) & $+$48:25:57.51(7)    \\
Spin frequency, $F$ ($\text{s}^{-1}$)\dotfill & 0.20365026489(1) &   5.4966921263(3) &  1.95238351856(1) \\ 
Spin frequency derivative, $\dot{F}$ ($\text{s}^{-2}$)\dotfill & $-6.14(1) \times 10^{-16}$ & $-9.1017(3) \times 10^{-13}$ & $-1.050(2) \times 10^{-15}$   \\
Spin period, $P$ (s)\dotfill & 4.9103790780(3) &  0.18192759882(1) & 0.512194448728(3)   \\
Spin period derivative, $\dot{P}$\dotfill & 1.480(4)$\times 10^{-14}$ &  3.01247(9)$\times 10^{-14}$ & 2.756(5)$\times 10^{-16}$  \\
Dispersion measure, DM ($\text{cm}^{-3}\,\text{pc}$)\dotfill & 331.3(5) & 104.35(8) & 85.7(3)   \\
\\
\hline
\textit{Fitting parameters} \\
First TOA (MJD)\dotfill & 58462.88 & 58259.105 & 58459.224    \\
Last TOA (MJD)\dotfill & 58956.12 & 58937.517  & 59083   \\
Timing epoch (MJD)\dotfill & 58650 & 58481.453266 & 58650.00   \\
Number of TOAs\dotfill & 74 & 211  & 23   \\
Total integration time (s)\dotfill & 80560 & 59081 & 78679  \\
Weighted RMS residual ($\mu\text{s}$)\dotfill &  2034.06 & 1521.69  & 119.79   \\
EFAC \dotfill & 2.09 & 2.04 & 0.927   \\
\\
\hline
\textit{Derived parameters} \\
Galactic longitude, $l$ ($^\circ$)\dotfill & 87.2061 & 81.1828  & 152.4276  \\
Galactic latitude, $b$ ($^\circ$)\dotfill & 1.6174 & 10.3473 & $-$3.1772   \\
DM distance, $d$ (kpc) \\
\multicolumn{1}{r}{NE2001} & 49.7 >  & 6.0 & 2.3   \\
\multicolumn{1}{r}{YMW16} & 8.9 & 9.0 & 1.8   \\
Characteristic age, $\tau_\text{c}$ (Myr)\dotfill & 5.4 & 0.098 & 30.0  \\
Surface magnetic field, $B_\text{surf}$ ($10^{10}\,\text{G}$)\dotfill & 862.6 & 236.9 & 38   \\
Spin-down luminosity, $\dot{E}$ ($10^{30}\,\text{erg}\,\text{s}^{-1}$)\dotfill & 4.9 & $1.9\cdot 10^5$ & 80.1   \\
\hline
\end{tabular}
\end{center}
\end{table*}

\begin{table}
\begin{center}
\caption{Timing solutions part 4.}\label{tab:timing4}
\begin{tabular}{ll}
\hline
Pulsar name & PSR J2338$+$4818    \\
\hline
\hline
\textit{Parameters} \\
Right ascension, $\alpha$ (J2000)\dotfill & 23:38:06.189(8) \\
Declination, $\delta$ (J2000)\dotfill & $+$48:18:32.19(7)  \\
Spin frequency, $F$ ($\text{s}^{-1}$)\dotfill & 8.42387236305(5)\\
Spin frequency derivative, $\dot{F}$ ($\text{s}^{-2}$)\dotfill &  $-$1.40(5)$\times 10^{-16}$ \\
Spin period, $P$ (s)\dotfill & 0.1187102506901(8)\\
Spin period derivative, $\dot{P}$\dotfill & 1.98$\times 10^{-18}$ \\
Dispersion measure, DM ($\text{cm}^{-3}\,\text{pc}$)\dotfill &  35.3(7) \\
Orbital period (days) \dotfill & 95.25536(2) \\
Eccentricity \dotfill & 0.0018237(9) \\
Projected semi-major axis (s) \dotfill & 117.58572(7)\\
Longitude periastron ($^\circ$) \dotfill & 99.65(2) \\
Epoch of periastron (MJD) \dotfill &  58868.444(7)\\
Binary model \dotfill & DD \\
\hline
\textit{Fitting parameters} \\
First TOA (MJD)\dotfill & 58462.941 \\
Last TOA (MJD)\dotfill &  59096.217 \\
Timing epoch (MJD)\dotfill & 58909.654 \\
Number of TOAs\dotfill & 63\\
Total integration time (s)\dotfill & 99263  \\
Weighted RMS residual ($\mu\text{s}$)\dotfill & 111.794\\
EFAC \dotfill &  0.76 \\
\\
\hline
\textit{Derived parameters} \\
Galactic longitude, $l$ ($^\circ$)\dotfill &  110.5436 \\
Galactic latitude, $b$ ($^\circ$)\dotfill & $-$12.7909 \\
DM distance, $d$ (kpc) \\
\multicolumn{1}{r}{NE2001} & 1.8\\
\multicolumn{1}{r}{YMW16} & 2.0\\
Characteristic age, $\tau_\text{c}$ (Myr)\dotfill & 950.11 \\
Surface magnetic field, $B_\text{surf}$ ($10^{10}\,\text{G}$)\dotfill & 1.55\\
Spin-down luminosity, $\dot{E}$ ($10^{30}\,\text{erg}\,\text{s}^{-1}$)\dotfill & 46.7 \\
\hline
\end{tabular}
\end{center}
\end{table}

\begin{table*}
\begin{center}
\caption{Pulse profile and polarisation properties of  FAST/EFF pulsars. The parameters listed for each pulsar from left to right are its pulse width at 50\,\% and 10\,\% of the profile peak ($W_{50}$ and $W_{10}$ respectively), RM is the rotation measure from \texttt{rmfit} and RM$_s$ is the value from the RM synthesis method, fraction of linear polarization ($L/I$), fraction of circular polarization ($V/I$), absolute circular polarisation ($\left|V\right|/I$) and the total integration time (T$_\text{obs}$) used to construct the integrated pulse profiles shown in Figure~\ref{fig:pulse_profiles}. 1$\sigma$ uncertainties are shown in parentheses.}\label{tab:pol_prof}
\begin{tabular}{lcccccccc}
\hline
PSR &$W_{50}$ & $W_{10}$  & RM & RM$_{s}$ & $L/I$ & $V/I$ & $\left|V\right|/I$ & T$_\text{obs}$ \\
  &  (ms) & (ms) & ($\text{rad}\,\text{m}^{-2}$) & ($\text{rad}\,\text{m}^{-2}$) & &  &   & (h) \\
 \hline
 J1822$+$2617 & 2.30 & 13.27 & 120(25) & 67(2) & 0.13(2) & $-$0.05(2) & 0.12(2) & 7200\\
 J2006$+$4058 & 11.20 & 19.00 & $<|200|$ & $<|200|$ & - & - & - & 14980\\
 J1942$+$3941 & 75.31 & 113.63 & $<|110|$ & $-$7(3)& 0.070(5) & $-$0.09(1) & 0.12(1) & 8500\\
 J2112$+$4058 & 43.61 & 107.05 & $<|160|$ & $-$7(2)& 0.15(1) & $-$0.07(1) & 0.10(1) & 3650\\
 J2129$+$4119 & 39.53 & 62.60 & $<|300|$ & $-$30(9)& 0.74(2) & $-$0.03(1) & 0.12(1) & 13740\\
 J1502$+$4653 & 29.08 & 41.06 & $<|120|$ & 60(8) & 0.080(2) & $-$0.04(1) & 0.05(2) &53970\\
 J2053$+$4718 & 57.50 & 86.25 & $-$792(71) & $-$785(1) & 0.10(1) & $-$0.08(2) & 0.04(2) & 14380\\
 J1951$+$4724 & 48.26 & 72.80 & $<|40|$ & 11(5) & 0.90(1) & $-$0.02(1) & 0.06(1) & 10770\\
 J0404$+$4831 & 73.38 & 87.04 & $-$111(70) & $-$89(9) & 0.48(3) & $-$0.08(3) & 0.11(3) & 3580\\
 J2337$+$4818 & 1.39 & 8.34 & - & - & 0.169(8) & +0.051(8) & 0.055(9) & 14290 \\
\hline
\end{tabular}
\end{center}
\end{table*}

\section{Discussion}\label{sec:discussion}
\subsection{J2338\texorpdfstring{$+$}\,4818: a mildly-recycled pulsar in the widest CO-WD binary system}
PSR J2338$+$4818 is located in the mildly recycled pulsar area of the $P-\dot{P}$ diagram (see Figure~\ref{fig:p-pdot}). The recycling scenario asserts that millisecond pulsars (MSPs) spun up  through accreting matter and angular momentum from a companion star (e.g. \citealt{Bhattacharya1991}).  During the transition to a MSP, a reduction in the magnetic field strength (B-strength) occurs (typically from $10^{12}$\,G to $10^{8}$\,G), either as a result of aging \citep{Goldreich1992} or accretion \citep{Bisnovatyi1974}. We describe a pulsar as mildly recycled when the accretion of matter from its companion is limited in time and did not spin the pulsar to rotational periods of the order of a few milliseconds, and its B-strength was not significantly reduced.

PSR J2338$+$4818 is in a wide binary of roughly 95.2 days, likely with a CO-WD companion. This conjecture is additionally supported by the pulsar's characteristic age of 0.95\,Gyr, which is consistent with the time needed to evolve a main-sequence (MS) star, with a mass between 3 and 6\, M$_{\odot}$, into a CO-WD (roughly between 0.3--0.6\, Gyr, see \citealt{Cruces2019}); and the fact that no optical counterpart was found in the catalogues within a 5' radius\footnote{SIMBAD Astronomical Database: http://simbad.u-strasbg.fr/simbad/}\footnote{ Gaia Archive at ESA: https://gea.esac.esa.int/archive/}. Considering PSR J2338$+$4818's distance of roughly 2\,kpc, if the companion was a MS star with a mass of roughly 1M$_{\odot}$ and an absolute magnitude of 4.8 (Sun's magnitude), the companion would have an apparent magnitude of 16 and thus likely visible by Gaia (among others).\\
Mildly recycled pulsars with CO-WD companions are thought to evolve from intermediate-mass (M>3 M$_{\odot}$ for the progenitor of the companion star) X-ray binaries (IMXBs). Furthermore, because the system is in a wide binary, the accretion phase likely underwent Case C \textit{Roche-lobe overflow} (RLO) and a common envelope phase \citep{Tauris2011,Tauris2012}. In this scenario, a high B-strength and discernible eccentricity are expected due to the inefficient mass transfer phase. According to the formalism presented in \citealt{Tauris2012}, to spin up a pulsar to a rotational period of 0.118 seconds, roughly 0.002M$_{\odot}$ needs to be accreted. This value is reasonable if we consider that the accretion occurs over a few megayears and that for IMXBs system up to $10^{-8}$ $\text{M}_{\odot}$  $\text{yr}^{-1}$ can be accreted by the pulsars based on the Eddington limit \citep{Tauris2012}.

\begin{figure}
    \centering
    \includegraphics[width=\linewidth, angle=0]{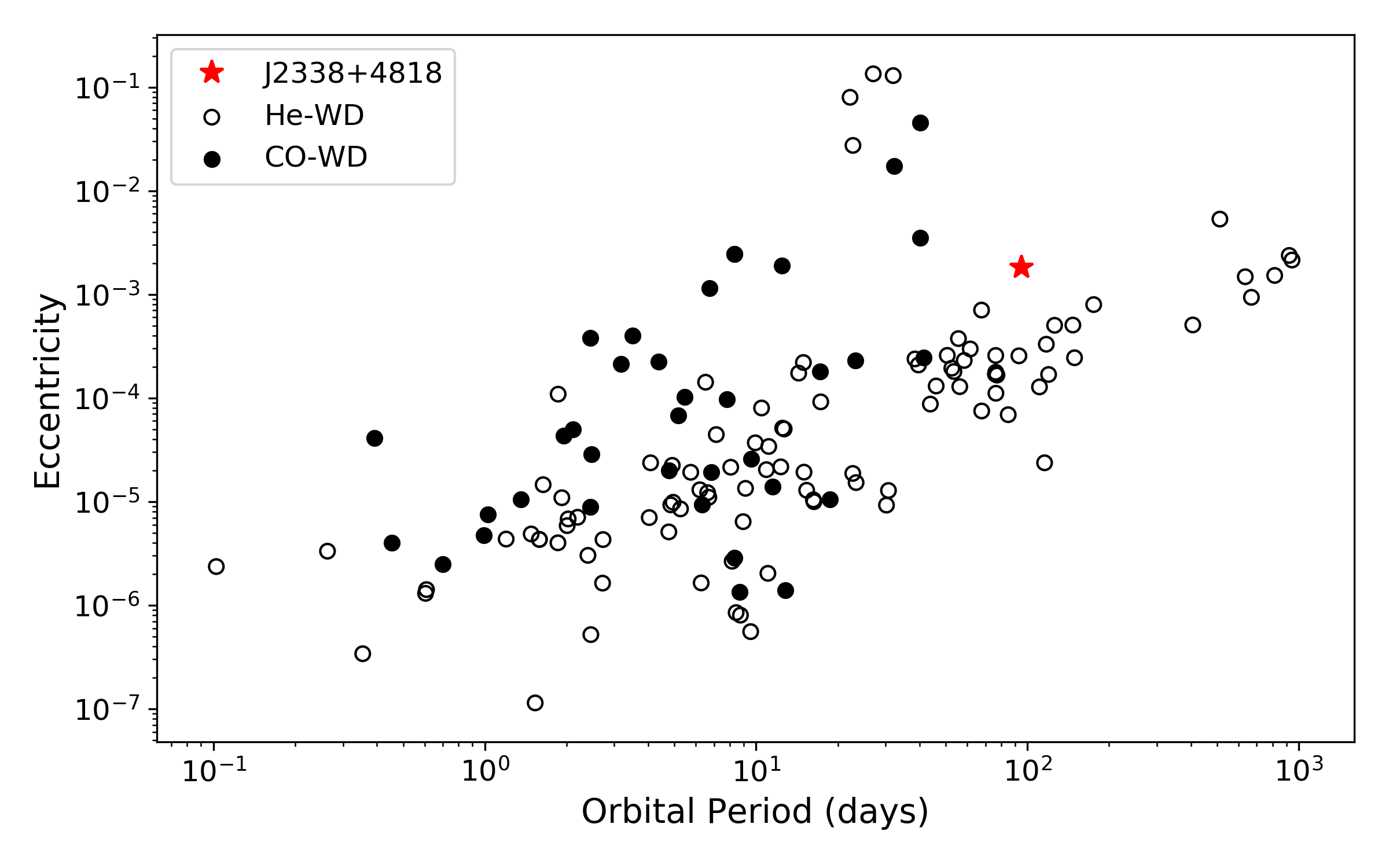}
    \caption{\textit{Orbital period - eccentricity} diagram of binary pulsars with either a He (\textit{black-open-circles}) or CO white dwarf companion (\textit{filled-black-circles}). We have excluded pulsars in globular clusters since their evolution involves companion exchanges or close interactions with other stars, and pulsars with B$>10^{11}$\,G as they have evolved through a different formation channel \citep{Tauris2000}.
    The orbital eccentricity of PSR~J2338$+$4818 lies in the trend observed for the other PSR - CO WD systems.}
    \label{fig:PbvsEcc}
\end{figure}

Figure~\ref{fig:PbvsEcc} shows the orbital period - eccentricity (P$_{b}$--e) relation of all the binary pulsars with either He-WD or CO-WD companions. We exclude pulsars in globular clusters since their evolution involves companion exchange, and pulsars with magnetic fields higher than B$>10^{11}$\,G, as young pulsars orbiting WDs are thought to have evolved from a different formation channel, where the close binary interaction and mass reversal led to a WD formed before the NS \citep{Tauris2000}. Thus, the systems shown in  Figure~\ref{fig:PbvsEcc} correspond to either fully or mildly recycled systems. We see that J2338$+$4818 follows the distribution of binaries, but most interestingly it is the widest pulsar CO-WD binary. Considering the estimated minimum mass for the CO-WD companion of 1.049 M$_{\odot}$, this also makes this system the widest WD binary with a companion mass over 0.8 M$_{\odot}$.

This pulsar is also interesting because of its potential long-term nulling. Currently, all the known long-term nulling pulsars are young sources \citep{Konar2019,Sheikh2021}, however, PSR J2338$+$4818 is an old mildly recycled pulsar. To show conclusively that PSR J2338$+$4818 is a long-term nuller, more observations are required.

\subsection{An old pulsar population}\label{subsec:old_pop}
To understand the population that the pulsars studied here trace we show a $P-\dot{P}$ diagram in Figure~\ref{fig:p-pdot} with four death lines, as described below. Each death line corresponds to the point at which the radio emission of a pulsar is expected to turn off as pair production can no longer sustain radio emission, according to different pulsar models. These lines should be taken with caution as the macro parameters describing the phenomena -- namely $P$ and $B$ -- are derived from a simplistic dipole model. Higher-order magnetic fields and micro physics described by the equation-of-state (EoS) might affect the pulsar's emission, and ultimately lead to a case by case turn off based on their mass. Instead, we should consider a death valley, where the likelihood of detecting a pulsar decreases the further below the death line a pulsar is (and the further into the valley).

The consideration of several death line models is motivated by the three pulsars from the FAST/EFF sample (PSRs J1942$+$3941, J2129$+$4119, and J1502$+$4653) that lie in the zone where only a handful of pulsars are expected. The models I, II, III, and IV shown in Figure~\ref{fig:p-pdot} are:
\begin{enumerate}[I)]
    \item Classic polar gap model, where the spin-down of the pulsar increases the thickness of the polar magnetospheric gap to a point where the minimum potential drop required for pair-production through curvature radiation is no longer supported \citep{Bhattacharya1992,Ruderman1975}. In the presence of a pure dipolar magnetic field, the death line is defined by:
    \begin{equation}
    \dfrac{B}{P^2}=0.17\times 10^{12}\,\rm{G/s^2}
    \end{equation}
    \item This model considers a more complex structure for the surface magnetic field, where the field lines are considerably more curved in comparison to the dipolar case, and hence the polar cap area is reduced \citep{Chen1993}. The death line is thus described by:
    \begin{equation}
        7\,\log\,(B)-13\,\log\,(P)=78 
    \end{equation}
    \item The polar gap model is corrected to account for the deviation from a flat spacetime when considering a general relativistic case \citep{Zhang2000}. This leads to a new death line described by:
    \begin{equation}
    \dfrac{9}{4}\,\log\,(P)-\log\,(\dot{P})=16.58
    \end{equation}
    \item The space-charge-limited flow (SCLF) model is an alternative scenario to the previously mentioned vacuum gap built above the NS surface. In the SCLF model, the charged particles are not bound to the surface and can be instead easily pulled from the surface. If additionally a curved space-time is considered, the result is a smaller potential drop needed for pair production \citep{Zhang2000}. The expression for the death line is:
    \begin{equation}
        2\,\log\,(P)-\log\,(\dot{P})=16.52
    \end{equation}
\end{enumerate}

Regarding the position of the FAST/EFF pulsars on the  $P-\dot{P}$ diagram in Figure~\ref{fig:p-pdot}, PSR~J1942$+$3941 -- located right at the death line from model I -- and  PSR J1502$+$4653 can  still be explained by the prediction from models II, III, and IV. However, PSR J2129$+$4119 is below all the death lines (models I, II, III, and IV). In Figure~\ref{fig:p-pdot} we also show the Parkes counterpart follow-up (\textit{non-filled black stars}). It is interesting to note that two of those sources lie near to the line for model I (although they remain above the model I death line).

We proceed hereon with analysis of the combined sample and refer to it as the FAST-UWB pulsars. As PSR J2338$+$4818 is a mildly recycled pulsar area, it is thus excluded.

The first noticeable pattern is that most of the pulsars seem to be located toward the right-hand-side of the normal pulsar zone, thus implying that they correspond to an older pulsar population. We compute the Kolmogorov-Smirnov statistic (KS test) for the age distribution of known pulsars \citep{Manchester2005} and the 21 FAST-UWB pulsars. Because MSPs are part of a different pulsar population, we exclude known pulsars with a $\dot{P}<10^{-18}$. The low $p-\text{value}=0.0015$ disfavors both samples being drawn from the same distribution.

One of the explanations for the population of older pulsars that FAST is finding is its sensitivity. This is consistent with the long integrations (>30 minutes) needed by Parkes and Effelsberg to carry out  follow-up observations. It is worth noting that all of the FAST-UWB pulsars were discovered through integrations at FAST of less than 52 seconds with the UWB receiver. The gain of 18 K/Jy for the new 19-beam receiver at FAST, promises to reveal a new population of pulsars with even lower luminosities, which may play a crucial role in understanding the emission physics of pulsars and the point where their emission turns off.

\subsection{Comparing the NE2001 and YMW16 electron density models}\label{subsec:ed_models}
\begin{figure*}
    \centering
    \includegraphics[width=\linewidth, angle=0]{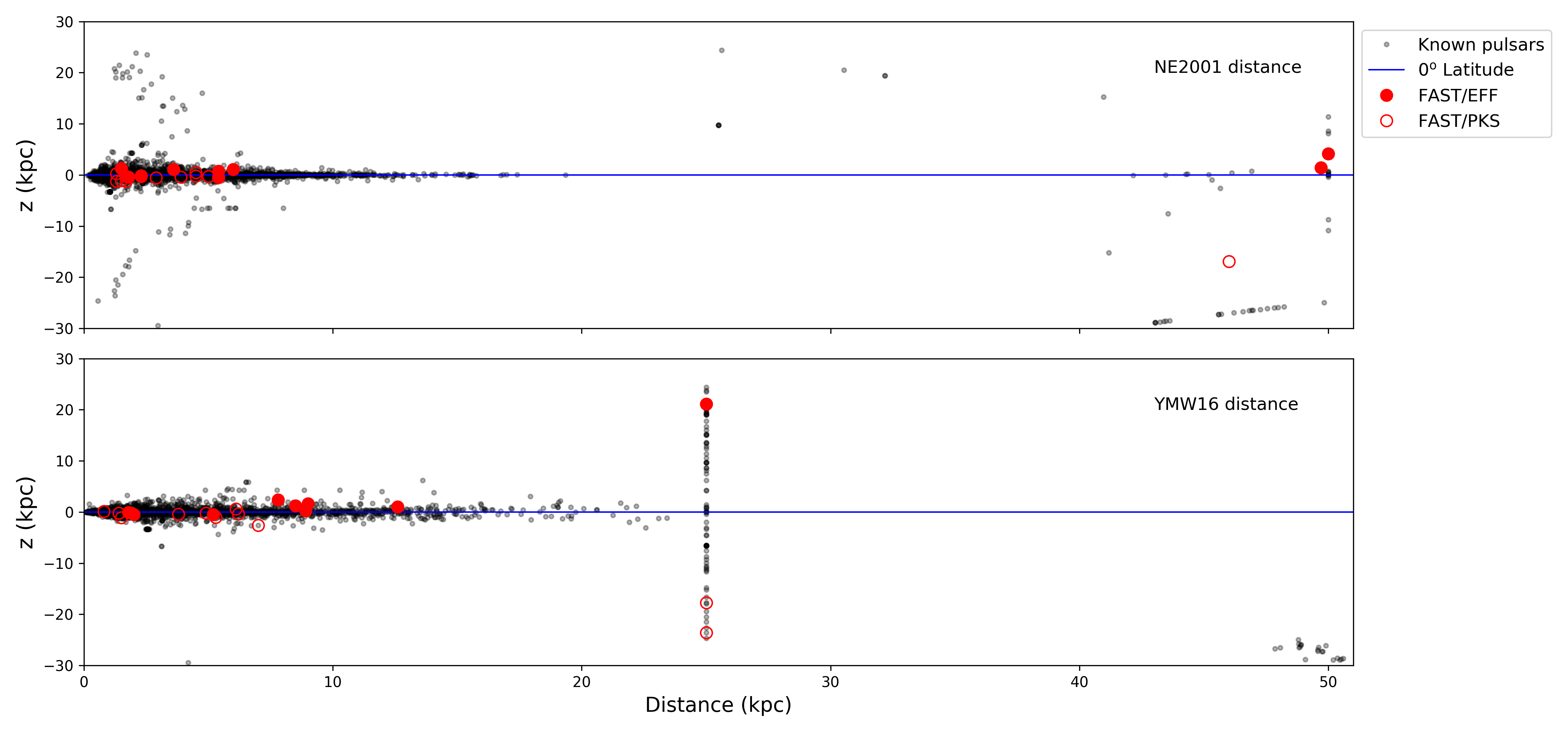}
    \caption{DM distance estimates based on the NE2001 (top) and YMW16 electron density model (bottom). The \textit{red-filled-circles} correspond to the FAST's UWB-receiver discoveries that were timed by Effelsberg and the \textit{open-red-circles} corresponds to the FAST pulsars timed by Parkes.}
    \label{fig:NE2001vsYMW20016}
\end{figure*}

We saw in Section~\ref{sec:results} that for three pulsars (PSRs J2006$+$4058, J1502$+$4653 and J2053$+$4718) both the NE2001 and the YMW16 model failed to correctly estimate the distance from the DM of the pulsar, and instead placed a lower limit on its value that is likely far greater than the true distance. Additionally, except for PSRs J2112$+$4119, J2129$+$4119, J0402$+$4825, and J2338$+$4818, the DM distance estimates from both models are not consistent. To better illustrate these differences we show in Figure~\ref{fig:NE2001vsYMW20016} the distance estimates derived from the NE2001 model (top) and from the YMW16 model (bottom) against the distance from the Galactic plane ($z$) based on each model. We include all the known pulsars and the FAST-UWB dataset.\\
First, we see that both models show an artifact due to the boundary of the electron density model. While for the YMW16 model the fiducial distance for out-of-the-Galaxy-pulsars appears at a value of 25\,kpc, for the NE2001 model the border of the model shows a more complex structure reached at 50\,kpc. Secondly, for the majority of the FAST-UWB pulsars the YMW16 distance estimation is larger than the NE2001 estimation for as much as a factor two. Pulsars at high $b$ (or $z$) such as PSR J1502$+$4653 seem to be more problematic for the YMW16 model than for the NE2001 model. While the YMW16 model places a lower bound of 25\,kpc on their distance, the NE2001 model locates them nearby, only within a couple of kilo-parsecs. On the other hand, the NE2001 model fails to estimate a distance for three sources located near the Galactic plane. For those pulsars, the YMW16 model derives a distance well before its boundaries.

To understand the difference in the derived distance estimation by the two models and the extent of their accuracy, it is compelling to recap on their assumptions. To map the distribution of free electrons in the Galaxy both models consider main large-scale components: a spiral-arm structure, the Galactic center component, an inner thin and outer thick disk component, and the local interstellar medium. The difference of both models lies in the electron densities ($n_e$) and extent of the before mentioned components, but most importantly how specific lines-of-sight are accounted for. For instance, the NE2001 model introduces artificial clumps and voids to account for the DM excess or lack towards the line-of-sight of particular pulsars, while the YMW16 model includes real features such as Nebulae, Local Bubble, Carina over-density, and Sagittarius under-density.

A key component to map the electron content is pulsars, due to the propagation effects that their radio pulses are subject to. Such effects are measured in the form of DM, scintillation bandwidth, and temporal and angular broadening due to scattering. These measures are only meaningful if they are accompanied by an independent distance estimation, for instance, through timing parallax or interferometric parallax, association to GCs or supernova remnants (SNR), or with HI absorption. Nonetheless, only $\sim$10\% of the known pulsars have independent distance estimations \citep{Manchester2005}. This low fraction of sources considerably limits the modeling of the free electron content.

The YMW16 is a later model of the free electron density, and made use of 189 independent pulsar distance estimations, in contrast to the 112 pulsars available at the time the NE2001 model was developed. The fact that only tens of pulsars -- out of the roughly 3000 known -- are located at high Galactic latitude makes their not-well-constrained DM derived distances not surprising. This is why the unconstrained DM of PSR J1502$+$465 according to YMW16 is expected. However, PSRs J2006$+$4058 and J2053$+$4718, the two FAST/EFF pulsars at the boundary of the NE2001 model, are located near the Galactic plane. For those cases, it is rather their Galactic Longitude ($\ang{77.1}$ and $\ang{87.2}$ respectively) what explains the unconstrained DM distances. At the time the NE2001 model was developed not many pulsars were known near the sky region between $\ang{70}<l<\ang{100}$. The under-representation of pulsars in that region in combination with the high DMs of 259.5 $\text{cm}^{-3}\,\text{pc}$ and 331.3 $\text{cm}^{-3}\,\text{pc}$ for J2006$+$4058 and J2053$+$4718, respectively, could explain why the model fails to estimate a distance and instead places a lower limit locating the pulsars outside of the Galaxy. We test this hypothesis by trying a series of DM values from 100 up to 400\,$\text{cm}^{-3}\,\text{pc}$ and coordinates in the above-mentioned region and with $|b|<\ang{6}$. We observed that for DM$>300$ $\text{cm}^{-3}\,\text{pc}$ (the exact value varies with $b$) the distance prediction was >50\,kpc.

For future updates of the electron density models, the current FAST pulsar discoveries have the potential to contribute to mapping along their line-of-sight, upon its independent distance determination.
\subsection{Detectability in previous surveys}
The FAST-UWB pulsars were found during FAST's early commissioning phase, with the use of a single-beam receiver in drift-scan. Despite a gain of 10 K/Jy -- in comparison with 18 \,K/Jy for the 19-beam receiver -- the UWB receiver provided a sensitivity comparable to Arecibo Observatory (AO).  Out of the more than 80 pulsars discovered by the FAST-UWB, 11 were reported in \citet{Cameron2020} and 10 in this work.

Even though the UWB receiver was equipped to cover a frequency range from 270 MHz to 1.62 GHz, the new pulsars were mostly found in the 270-500 MHz band due to the combination of the drop in antenna efficiency for frequencies above 800 MHz and the drop in sky temperature, which translates into a loss of relative advantage of FAST since UWB was uncooled. Because of this, we compare with other pulsar surveys carried out in the low-frequency regime (< 500\,MHz) and explore whether the FAST-UWB pulsars could have been discovered earlier. We identify four main surveys: AO327 carried by AO at a central frequency of 327 MHz \citep{Deneva2013}; GBT350 \citep{Hessels2008} and GBNCC \citep{Stovall2013}, both carried with the Green Bank Telescope (GBT) at a central frequency of 350 MHz; and LOFAR's LOTAAS survey at a central frequency of 135 MHz \citep{Sanidas2019}. We summarize in Table~\ref{tab:surveys} the above-mentioned surveys and the main parameters used in the sensitivity comparison. We explore exclusively the detectability of the FAST-UWB sample - composed of the 11 pulsars reported by \citet{Cameron2020} and 10 from this work. For the detectability of the FAST-PKS sources with surveys carried at L-band refer to the discussion in \citet{Cameron2020}.

The flux density for the FAST/PKS pulsars at 1.4 GHz ($S_{1400}$) ranges from roughly 0.1 to 1.0 mJy \citep{Cameron2020}. Because we have not performed flux calibration of the FAST/EFF pulsars, we restrict ourselves to the flux ranges provided by the FAST/PKS sample. As they trace the same population, this assumption is reasonable. We scale the flux density to a frequency of 300 MHz with the assumption of a spectral index $\alpha=-1.6$ \citep{Bailes2008} and find that the flux density ranges from 1.1 to 11.7 mJy.

Considering the LOTAAS survey sensitivity to pulsars with flux densities ($S$) above 5 mJy for a 3600\,s integration, and their visibility to declinations above $\ang{0}$ (see Table~\ref{tab:surveys}), the FAST/EFF pulsars as well as PSRs J1945$+$1211 and J1919$+$2621 from the FAST/PKS sample could have been detected. Furthermore, because of the long integration of 3600\,s, scintillation would not prevent a detection. The GBT350 survey, due to their northern Galactic plane restriction to declination above $\ang{38}$, could have potentially detected all the FAST/EFF pulsars except for PSR J1822$+$2617. In contrast, the GBNCC survey carried as well by GBT, explored all their observable sky (declination above $-\ang{40}$). GBNCC sensitivity to pulsar with $S>0.7$\,mJy could have potentially detected all of the FAST-UWB pulsars. However, scintillation could have played a role for the low DM pulsar given the short integrations of roughly two minutes. For the pulsar with high DM, pulse broadening due to scattering is likely not the reason, due to the long spin periods of the pulsars and the expected broadening of a couple of milliseconds at 300\,MHz.

AO327 survey is the most comparable survey to the FAST pulsar survey reported here. With a sensitivity to pulsars with $S>0.3$\,mJy, and a comparable integration to FAST -- at the testing phase -- AO327 could have seen all the FAST/EFF pulsars and PSRs J1945$+$1211, J2323$+$1214, and J1919$+$2621 from the FAST/PKS sample. Again, scintillation could be invoked for the low DM pulsars, but once more, pulse broadening is likely not the reason for the high DM sources.\\
The determination of FAST-UWB pulsar's flux densities at low frequencies (such as $S_{400}$) could provide insights on why these pulsars were not detected before. We suggest to explore the data of the corresponding surveys by folding the observations (if existing) with the ephemeris provided in \citet{Cameron2020} and in this work.
\begin{table*}
\caption{Pulsar surveys at low frequencies and their parameters. The first column list the survey name, the second column names the telescope (abbreviations: EFF=Effelsberg, GBT=Green Bank, AO=Arecibo observatory), $f$ corresponds to the central observing frequency and $\Delta\,f$ corresponds to the frequency bandwidth, $\text{T}_{\text{obs}}$ is the observation time needed to achieved a minimum observable flux $S_{\text{min}}$.}\label{tab:surveys}
\begin{tabular}{lccccccc}
\hline
\hline
Survey & Telescope & $f$ & $\Delta\,f$ & T$_{\text{obs}}$ & Gain & S$_{\text{min}}$ & Visibility \\
& & (MHz) & (MHz)& (s) & (K/Jy) & (mJy)\\
\hline
AO327 & AO & 327 & 69 & 60 & 11.0 & 0.3 & DEC $-\ang{1}$--$+\ang{38}$\\
GBT350 & GBT & 350 & 50 & 149 & 2.0 & 1.0 & DEC $>+\ang{38}$ \\
GBNCC & GBT & 350 & 100 & 120 & 2.0 & 0.7 & DEC $>\ang{-40}$\\
LOTAAS & LOFAR & 135 & 32 & 3600 & 1.7 & 5.0 & DEC $>\ang{0}$\\
FAST-UWB & FAST & 500 & 530 & 52 & 9.0 & 0.2 & DEC $\ang{-14}$--$+\ang{66}$\\
\hline 
\end{tabular}
\end{table*}

\section{Conclusions}\label{sec:conclusions}
We reported 10 new pulsars from the pilot runs of the CRAFTS survey. These pulsars were discovered by the FAST-UWB and confirmed and timed by the Effelsberg telescope. Our main findings are:

1) Most FAST-UWB discoveries are located in the normal pulsar zone on the $P-\dot{P}$ diagram and seem to trace an older population than the general normal pulsars based on a KS test.

2) PSRs J2053$+$4718 and J0402$+$4825 are located at low latitudes ($b<|\ang{3.5}|$), near the Galactic plane; PSR J1502$+$4653 at high Galactic latitude; the remaining FAST/EFF sources at mid-Galactic latitude.

3) Notable sources: PSRs J1951$+$4724 is a young and energetic pulsar; J2129$+$4119, J1942$+$3941, and J1502$+$4654 are old pulsars below the classic death lines; and perhaps the most interesting pulsar in our sample, PSR J2338$+$4818 which is a pulsar in a binary orbit.
PSR J2338$+$4818 turns out be a mildly recycled pulsar with a massive CO-WD companion, and in the widest (95.2 days) orbit among such binaries. Assuming a pulsar mass of 1.4\,M$_{\odot}$, the companion is to have a minimum mass of 1.049 M$_{\odot}$. Such a system likely evolved from an IMXB, where the inefficient mass transfer through type C Roche-lobe overflow phase led to a pulsar that is not fully spun-up (P=0.1187 s) and in an orbit with discernible eccentricity (0.0018).
J2338$+$4818 is also the widest binary with a massive white dwarf companion (M$>0.8\>\,$M$_{\odot}$), among the recycled/mildly-recycled systems. J2338$+$4818 exhibits turn-offs on time scales longer than $\sim$ 1 hour.

4) High degrees ($>70\%$) of linear polarization were found in PSRs J2129$+$4058 and J1951$+$4724. PSRs J2112$+$4058, J2129$+$4119, J1951$+$4724, and J0402$+$4825 had well-measured PA swing for constraining the magnetic inclination angles. Only a small fraction ($<12\%$) of circular polarization was detected for all the pulsars, except for PSR J2006$+$4058.

5) Conspicuous discrepancies were found between two commonly used electron density models, NE2001 and YMW16. For PSRs J2006$+$4058 (DM=259.5 $\text{cm}^{-3}\,\text{pc}$) and J2053$+$4718  (DM= 331.3 $\text{cm}^{-3}\,\text{pc}$), the NE2001 places a lower limit of 50\,kpc on their distance, locating them outside the Galaxy. The YMW16 model estimates are 12.6 and 8.0\,kpc, respectively. YMW16 model fails to estimate a distance to PSR J1502$+$4653 despite its low DM of 26.6 $\text{cm}^{-3}\,\text{pc}$ and instead yields a lower limit of 25\,kpc. Since these pulsars are unlikely to be extra-galactic, we may attribute these oddities to poor sampling in the electron density models at high Galactic latitudes and for longitudes between $\ang{70}<l<\ang{100}$. 

Finally, we stress the importance of collaboration in the era of radio astronomy with most sensitive radio telescopes such as FAST, Meerkat, and SKA. The high number of discoveries expected for each of these telescopes is hard if not impossible to follow-up by a single facility. Allocating timing of the brighter sources to telescopes such as SRT, GMRT, VLA, Parkes, Lovell, GBT, and Effelsberg, can enhance the discovery rate as well as facilitate proper follow-ups.

\section*{Acknowledgements}
We thank the referee for helpful comments that have contributed to improve this manuscript. This work is funded by the National Natural Science Foundation of China under Grant No. 11988101, 11690024, 11725313, 11743002, 12041303, and 11873067, the CAS-MPG LEGACY project, the  National Key R\&D Program of China No. 2017YFA0402600, the CAS Strategic Priority Research Program No. XDB23000000, the National SKA Program of China No. 2020SKA0120200.   
This publication is based on observations with the 100-m telescope of the Max-Planck-Institut fuer Radioastronomie at Effelsberg. We thank Dr. A. Kraus for scheduling our observations. We thank Dr. K. Lackeos, Dr. N. Porayko, J. Wongphecauxon and Dr. R. Main for helpful discussions. Lei Qian is supported by the Youth Innovation Promotion Association of CAS (id.~2018075). This work made use of data from the Five-hundred-meter Aperture Spherical radio Telescope (FAST), a Chinese national mega-science facility, built and operated by the National Astronomical Observatories, Chinese Academy of Sciences (NAOC). We appreciate the efforts of all those in the FAST Collaboration for their support and assistance during these observations. 

\section*{Data availability}
The data underlying this article will be shared on reasonable request to the corresponding author.

\bsp	
\label{lastpage}

\bibliographystyle{mnras}
\bibliography{bibliography}
\end{document}